\documentclass[11pt,a4paper]{article}
\usepackage{jheppub}
\usepackage{graphicx}
\usepackage{extarrows}
\usepackage{amsmath}
\usepackage{mathtools}
\input epsf
\usepackage{epsfig}
\usepackage{tikz}
\usepackage{subfigure,tikz}
\usetikzlibrary{backgrounds,automata}
\usepackage{amssymb}
\usepackage{graphics}
\usepackage{subfigure}
\usepackage[active]{srcltx}
\usepackage{amsthm}
\usepackage[T1]{fontenc} 
\usepackage{graphicx}
\usepackage{upgreek}
\usepackage{breqn}
\usepackage{booktabs}

\newcommand{\bea}{\begin{eqnarray}}
\newcommand{\eea}{\end{eqnarray}}
\newcommand{\bean}{\begin{eqnarray*}}
\newcommand{\eean}{\end{eqnarray*}}
\newcommand{\nn}{\nonumber \\}

\def\no{\nonumber}

\def\WH #1{\widehat{#1}}

\def\co{\,,}
\def\ed{\,.}

\def\abs#1{\left| #1\right|}

\def\eref#1{(\ref{#1})}

\def\d{{\rm d}}

\def\a{{\alpha}}

\def\d{\partial}

\def\eps{\epsilon}

\def\Label#1{\label{#1}%
  \smash{\hbox to0pt{\raise1ex\hbox{\tiny[#1]}\hss}}}

\allowdisplaybreaks[4]
\newcommand{\longline}[2]{
  \begin{gathered}
   #1 = 
      \begin{minipage}[t]{.7\displaywidth}
      \raggedright\linespread{1.2}\selectfont
      \begin{math}
    #2      
      \end{math}
      \end{minipage}
    \end{gathered}
}
\newcommand{\longlongline}[2]{
  \begin{aligned}
    &#1 = \\
    &~~\begin{minipage}[t]{.95\displaywidth}
        \raggedright\linespread{1.2}\selectfont
          \begin{math}            
            #2      
          \end{math}
      \end{minipage}
  \end{aligned}
}

\title{\boldmath Reduction of General One-loop Integrals Using Auxiliary Vector}

\author[a,b,c,d]{Bo Feng,}
\author[a]{Tingfei Li,}
\author[a]{Hongbin Wang,}
\author[a]{Yaobo Zhang}

\affiliation[a]{Zhejiang Institute of Modern Physics, Zhejiang University, Hangzhou, 310027, P. R. China }
\affiliation[b]{Beijing Computational Science Research Center, Beijing 100084, China}
\affiliation[c]{Center of Mathematical Science, Zhejiang University, Hangzhou, 310027, P. R. China}
\affiliation[d]{Peng Huanwu Center for Fundamental Theory, Hefei, Anhui 230026, China}

\emailAdd{fengbo@zju.edu.cn}
\emailAdd{tfli@zju.edu.cn}
\emailAdd{21836003@zju.edu.cn}
\emailAdd{yaobozhang@zju.edu.cn}

\abstract{ 
	As a key method to deal with loop integrals, Integration-By-Parts (IBP) method can be used to do reduction as well as establish the differential equations for master integrals. However, when talking about tensor reduction, the Passarino-Veltman (PV) reduction method is also widely used for one-loop integrals. Recently, we have proposed an improved PV reduction method, i.e., the PV reduction method with auxiliary vector $R$, which can easily give analytical reduction results for any tensor rank. However, our results are only for integrals with propagators with power one. In this paper, we generalize our method to one-loop integrals with general tensor structures and propagators with general powers. Our ideas are simple. We solve the generalised reduction problem by combining differentiation over masses and proper limit of reduction with power-one propagators. Finally, we demonstrate our method with several examples. With the result in this paper, we have shown that our improved PV-reduction method with auxiliary vector is a self-completed reduction method for one-loop integrals.   
}


\begin{document}
\maketitle
\flushbottom
\section{Introduction}

For loop integrals, the most used strategy is the reduction method. For one-loop integrals, the unitarity cut method \cite{Bern:1994zx,Bern:1994cg,Britto:2004nc,Britto:2005ha,Britto:2006sj,Anastasiou:2006jv,Britto:2006fc,Anastasiou:2006gt,Britto:2007tt} has been proved to be an efficient reduction method. 
However, there are still some remaining problems for the unitarity cut method. The first one is how to do the unitarity cut when propagators have higher powers. It is solved in \cite{Feng:2021spv} where the technique of differentiation over auxiliary masses has been used. The second problem is that the well established unitarity cut method can not find the reduction coefficients of tadpoles. Some efforts have been taken by adding auxiliary propagator in \cite{Britto:2009wz} or using the single cut in \cite{Britto:2010um}.
However, the separation of tadpole coefficients and other contributions is not straightforward in \cite{Britto:2009wz,Britto:2010um}. The problem has been reconsidered   in \cite{Feng:2021enk} using the familiar idea of PV-reduction \cite{Passarino:1978jh} with an improvement, i.e., adding an auxiliary vector $R$.
Later, we find that one can obtain reduction coefficients of all other master integrals except tadpoles with the method, as shown in \cite{Hu:2021nia}.
In these two papers \cite{Feng:2021enk,Hu:2021nia} for the PV tensor reduction, the powers of all propagators are fixed to be just one.
However, when we state a reduction method is complete, it should tell us how to reduce integrals with general tensor structures and propagators with arbitrary powers. 
In this paper, we will show how to use our newly developed method, the {\bf improved PV-reduction method with auxiliary vectors}, to reduce one-loop integrals having propagators with general powers. 
Combining the results for arbitrary tensor structures in \cite{Feng:2021enk,Hu:2021nia}, we can claim our method is a {\sl complete one-loop integral reduction method}.
For a general one-loop integral, i.e., with general tensor structure and  arbitrary power of propagators, we can use a trick as shown in  \cite{Feng:2021spv}, i.e.,
\bea  &&{\cal M}[ {\cal N}, \{a_1,...,a_n\}] \equiv   \int {d^{D}\ell \over (2\pi)^{D/2}} { {\cal N}[\ell]\over \prod_{j=1}^{N} ((\ell-K_j)^2- m_j^2+i\eps)^{a_i}}\nn
& = & \left\{\prod_{j=1}^{N} {1\over (a_j-1)!}{d^{a_j-1}\over d \eta_j^{a_j-1}}\int {d^{D}\ell \over (2\pi)^{D/2}} { {\cal N}[\ell]\over \prod_{j=1}^{N} ((\ell-K_j)^2- m_j^2-\eta_j+ i\eps)}\right\}\Bigg\vert_{\eta_j\to 0}
~~~~\label{gen-1} \eea
to change the problem to the tensor reduction  with power one for all propagators, thus we get
\bea {\cal M}[ {\cal N}, \{a_1,...,a_n\}]=\left\{\prod_{j=1}^{N} {1\over (a_j-1)!}{d^{a_j-1}\over d \eta_j^{a_j-1}} \sum_{i} c_i(\eta_i) {\cal I}_i (\eta_i)\right\}\Bigg\vert_{\eta_j\to 0}
~~~~\label{gen-2}\eea
where all $c_i(\eta_i)$ can be found using the framework given in  \cite{Feng:2021enk,Hu:2021nia}. The action of ${d\over d \eta_j}$
over the scalar basis ${\cal I}_i (\eta_i)$ will
produce scalar integrals  with propagators having general powers. Thus to give the reduction of general one-loop integrals,
we only need  to know the reduction of scalar basis  with propagators having general powers\footnote{
This problem has been partially addressed in \cite{Feng:2021spv} using the unitarity cut method. However, since the tadpole coefficients are still missing by the unitarity cut method, we reconsider the problem in this paper.}.
Moreover, as explained in this paper (see also  \cite{Feng:2021spv}), there are recursion relations and symmetric relations among scalar integrals with the general power of propagators.
Thus all tasks are reduced to the reduction of scalar integrals with only one propagator having power two.

With the above explanations,  in this paper, we will show how to solve the problem by taking proper limits of the tensor reduction of integrals with all propagators having power one. The connection between tensor reduction and general scalar reduction is not so surprising since such a phenomenon also appears in the familiar IBP relations \cite{Tkachov:1981wb,Chetyrkin:1981qh}. Our proper limit approach is just another prospect to see the same thing.

The plan of the paper is following. In section two, we will explain our idea carefully and set up the general frame. In section three, we will
use our framework to calculate the scalar reduction with one propagator having power two for bubble and triangle topologies. 
In section four, we give the conclusion and some discussions.
In the appendix, we present partial results for box and pentagon, and the full results are collected in an attached \href{https://github.com/Wanghongbin123/HigherPole}{Mathematica file}.

\section{Set up}
In this section, we will explain our idea to get the reduction of general scalar one-loop integrals. Before doing so, let us set up our notations. After introducing the auxiliary vector $R$, the general tensor one-loop integral with high poles can be written as
\begin{equation}
	I^{(m)}_{\mathbf{a}_n}\equiv I^{(m)}_{a_0,a_1,a_2,\cdots,a_n}=\int {d^{D}\ell\over (2\pi)^{D}} { (2\ell\cdot R)^m\over D_0^{a_0}D_1^{a_1}\cdots D_n^{a_n}},~\label{Tensor-High-Pole}
\end{equation}
where 
\begin{align}
D_0\equiv\ell^2-M_0^2,~~D_i\equiv (\ell-K_i)^2-M_i^2,i=1,2,...,n\ed
\end{align}
We denote a $(n+1)$-component vector $\mathbf{a}_n\equiv \{a_0,a_1,\cdots,a_n\}$ with any element being non-negative integer and $\abs{\mathbf{a}_n}\equiv \sum_{s=0}^{N}a_s>0$. It is well-known that any  $D=(d-2\eps)$-dimensional one-loop integral can be reduced to master
scalar integrals\footnote{ The conclusion is true for any $D$ dimensional space-time. For  $D=4-2\epsilon$, the topology of basis
	is from tadpoles to pentagons. For this reason, we will do examples up to pentagon. }
\begin{equation}
	I^{(m)}_{\mathbf{a}_n}=\sum^{d+1}_{ \substack{\abs{\mathbf{b}_n}=1,\\ b_i=\{0,1\}}}C^{(m)}_{\mathbf{a}_n\to \mathbf{b}_n}I_{\mathbf{b}_n}
	~~~\label{red-1}
\end{equation}
where $C^{(m)}_{\mathbf{a}_n\to \mathbf{b}_n}$ are the reduction coefficients. 
When all $a_i$ are one or zero, \cite{Feng:2021enk,Hu:2021nia} have shown how to efficiently find reduction coefficients using  two types of differential operators $\mathcal{D}_i$ and $\mathcal{T}$, which are   defined as
\begin{equation}
	\mathcal{D}_i\equiv K_i\cdot {\d \over \d R},~~i=1,...,n;~~~~ \quad \mathcal{T}\equiv {g}^{\mu\nu}{\d\over \d R^\mu}
	{\d \over \d R^\nu}\ed
\end{equation}

To have an idea of getting the reduction coefficients for integrals having propagators with  general powers, let us consider the simplest example, i.e., the reduction of tadpole topology
\bea \int {d^{D}\ell\over (2\pi)^{D}} {1\over (\ell^2-m^2)^2}=\a \int {d^{D}\ell\over (2\pi)^{D}} {1\over (\ell^2-m^2)}\ed~~~\label{tad-1} \eea
Aimed to find the coefficient $\a$, we do the following. At one side, taking the derivative over the $m^2$ to the known tensor  reduction result
\bea  \int {d^{D}\ell\over (2\pi)^{D}} {(2\ell\cdot R)^2\over (\ell^2-m^2)}=b \int {d^{D}\ell\over (2\pi)^{D}} {1\over (\ell^2-m^2)}\co~~~\label{tad-2} \eea
we get
\bea \int {d^{D}\ell\over (2\pi)^{D}}  {(2\ell\cdot R)^2\over (\ell^2-m^2)^2}=
{\d b\over \d m^2} \int {d^{D}\ell\over (2\pi)^{D}} {1\over (\ell^2-m^2)}+b \int {d^{D}\ell\over (2\pi)^{D}} {1\over (\ell^2-m^2)^2}\ed~~~\label{tad-3} \eea
At another side,  starting with the known tensor bubble reduction
\bea & & \int {d^{D}\ell\over (2\pi)^{D}} {(2\ell\cdot R)^2\over (\ell^2-M_0^2)((\ell-K)^2-M_1^2)} \nn
& = & c_2 \int {d^{D}\ell\over (2\pi)^{D}} {1\over (\ell^2-M_0^2)((\ell-K)^2-M_1^2)}+ c_{1;1} \int {d^{D}\ell\over (2\pi)^{D}} {1\over (\ell^2-M_0^2)}\nn
& & +
c_{1;2} \int {d^{D}\ell\over (2\pi)^{D}} {1\over ((\ell-K)^2-M_1^2)}~,~~\label{tad-4}
\eea
after taking the limits $M_0,M_1\to m$ and $K\to 0$, we get
\bea & & \int {d^{D}\ell\over (2\pi)^{D}} {(2\ell\cdot R)^2\over (\ell^2-M_0^2)^2}= \WH c_2 \int {d^{D}\ell\over (2\pi)^{D}} {1\over (\ell^2-m^2)^2}+ (\WH c_{1;1}+\WH c_{1,2}) \int d^{D}\ell {1\over (\ell^2-m^2)}~~~\label{tad-5}
\eea
where $\WH c$ is to emphasize the expressions after taking the limit. Comparing both sides, we get relation
\bea & & (b-\WH c_{2}) \int {d^{D}\ell\over (2\pi)^{D}} {1\over (\ell^2-m^2)^2}= (\WH c_{1;1}+\WH c_{1,2}-{\d b\over \d m^2}) \int {d^{D}\ell\over (2\pi)^{D}} {1\over (\ell^2-m^2)}\ed~~~\label{tad-6} \eea
With the explicit expressions of various coefficients, under the limit $M_0,M_1\to m$ and $K\to t R$ with $t\to 0$ we have

\bea 
  c_{1;1}&= & \frac{\left(s_{00} s_{11}-D s_{01}^2\right) f_1}{(D-1) s_{11}^2}\to -s_{00}\nn
  c_{1;2}&= & \frac{s_{01}^2 \left((3 D-4) s_{11}+D M_0^2-D M_1^2\right)+s_{00} s_{11} \left(s_{11}-M_0^2+M_1^2\right)}{(D-1) s_{11}^2}\to 3 s_{00}\nn
  c_2&= &\Bigg[\frac{s_{01}^2 \left(-2 M_0^2 \left(D M_1^2-(D-2) s_{11}\right)+D \left(M_1^2-s_{11}\right){}^2+D M_0^4\right)}{(D-1) s_{11}^2}\nn &&-\frac{\left(M_0^2-2 M_1 M_0+M_1^2-s_{11}\right) \left(M_0^2+2 M_1 M_0+M_1^2-s_{11}\right) s_{00}}{(D-1) s_{11}}\Bigg]\to 0 \nn
  b&= & {4s_{00}m^2\over D},~~~{\d b\over \d m^2}={4s_{00}\over D}
  ~~~\label{tad-7}
\eea
where $f_1 \equiv K^2-M_0^2+M_1^2$, $s_{00} \equiv R^2$, $s_{01} \equiv R\cdot K$, $s_{11}  \equiv K^2$ and the "$\to$" means the results after taking limits, so finally we have
\bea  \int {d^{D}\ell\over (2\pi)^{D}} {1\over (\ell^2-m^2)^2}={(D-2)\over 2m^2}\int {d^{D}\ell\over (2\pi)^{D}} {1\over (\ell^2-m^2)}\ed~~~\label{tad-8} \eea
Since our results should be independent of $R$, we can take $R$ properly. For example, $R\cdot K=s_{01}=0$. Under this gauge choice and with the limit $f_1\to s_{11}= K^2$, we can  get the same results as \eref{tad-8} with another limit choice:
\bea 
  \WH c_{1;1}={s_{00}\over D-1},~~~\WH c_{1;2} = {s_{00}\over D-1},~~~~\WH c_2= {4 M_0^2 s_{00}\over D-1},~~~
  {(\WH c_{1;1}+\WH c_{1,2}-{\d b\over \d m^2})\over (b-\WH c_{2})}={D-2\over 2m^2}\ed
\eea 
Although with different choices of limit processes, the final result is the same.

The above procedure can be written more abstractly using the notation in \eref{Tensor-High-Pole} as follows. 
Starting from the identity
\begin{equation}
	I_{2}^{(m)}=\lim_{D_{1}\to D_0}I^{(m)}_{1,1}=\partial_{M_0^2} I_{1}^{(m)}~,
  \label{Eqtadpole}
\end{equation}
when expanding the middle term of \eqref{Eqtadpole} we could get
\begin{equation}
  \begin{aligned}
  \lim_{D_{1}\to D_0}I^{(m)}_{1,1} &= \lim_{D_{1}\to D_0}
  C_{1,1 \rightarrow {1,1}}^{(m)} I_{1,1}
  +
  \lim_{D_{1}\to D_0}
  \left(C_{1,1 \rightarrow {0,1}}^{(m)} I_{0,1}
  +C_{1,1 \rightarrow {1,0}}^{(m)} I_{1,0}
  \right)\\
  &=\left(\lim_{D_{1}\to D_0}
  C_{1,1 \rightarrow 1,1}^{(m)}\right) I_{2}
  +
  \left(\lim_{D_{1}\to D_0}
  \left(C_{1,1 \rightarrow {0,1}}^{(m)}
  +C_{1,1\rightarrow {1,0}}^{(m)}\right)
  \right) I_{1}\co
\end{aligned}
\label{EqtadpoleA}
\end{equation}
while differentiating each part of the rightmost term in  \eqref{Eqtadpole} we get
\bea
  \partial_{M_0^2} I_{1}^{(m)}&= &
  \partial_{M_0^2} \left( C_{1 \rightarrow 1}^{(m)} I_1\right)=
  \left(\partial_{M_0^2}  C_{1 \rightarrow 1}^{(m)}\right) I_1
  +  C_{1 \rightarrow 1}^{(m)} I_{2}\ed
  ~~~\label{EqtadpoleB}
\eea
Identifying the two equations \eqref{EqtadpoleA} and \eqref{EqtadpoleB}, we solve $I_{2}$ as
\bea I_2 & = & {\left(\lim_{D_{1}\to D_0}
	\left(C_{1,1 \rightarrow {0,1}}^{(m)}
	+C_{1,1 \rightarrow {1,0}}^{(m)}\right)
	\right)-\left(\partial_{M_0^2}  C_{1 \rightarrow 1}^{(m)}\right) \over C_{1 \rightarrow 1}^{(m)}-\left(\lim_{D_{1}\to D_0}
	C_{1,1 \rightarrow 1,1}^{(m)}\right)}I_1\ed\eea
Of course, we need to ensure that the denominator is nonzero; thus, the tensor rank $m$ should be appropriately chosen.

\subsection{The framework}

Having above warm up example, now we set up the general framework of getting the reduction of integrals with propagators having higher powers.
Our idea is to consider the identity
\begin{equation}
	I^{(m)}_{a_0,a_1,\cdots,a_r+1,\cdots,a_n}=\lim_{D_{n+1}\to D_r}I^{(m)}_{a_0,a_1,a_2,\cdots,a_n,1}={1\over a_r}{\partial \over \partial {M_r^2}}I^{(m)}_{a_0,a_1,\cdots,a_r,\cdots,a_n}~~\label{Basic-Relation}
\end{equation}
where $a_r\not=0,r=0,1,2,\cdots,n$. It is worth to point out that using ${\partial \over \partial {M_r^2}}$ to increase
the power of a given propagator is a well known trick and has been used  many times by many people. We repeat the
discussion here is for self-completeness and the aim is to show that we can reduce the problem to the case one and only
one propagator has power two, while others, power one.

To simplify out notations, we define the  action $\mathbf{r}^\pm$  on the  vector $\mathbf{a}_n$ with $a_r>0$ as
\begin{equation}
	\mathbf{r}^+\mathbf{a}_n=\{a_0,a_1,\cdots,a_r+1,\cdots,a_n\};~~\mathbf{r}^-\mathbf{a}_n=\{a_0,a_1,\cdots,a_r-1,\cdots,a_n\}.
\end{equation}
Then \eref{Basic-Relation} can be written as
\begin{equation}
	I^{(m)}_{\mathbf{r}^+\mathbf{a}_n}=\lim_{D_{n+1}\to D_r}I^{(m)}_{\mathbf{a}_n,1}={1\over a_r}{\partial \over \partial {M_r^2}}I^{(m)}_{\mathbf{a}_n}\ed~~\label{Basic-Relation-2}
\end{equation}
By the reduction assumption, the expansion of $I^{(m)}_{\mathbf{a}_n,1}$ and $I^{(m)}_{\mathbf{a}_n}$ are known, while the expansion of $I^{(m)}_{\mathbf{r}^+\mathbf{a}_n}$
\begin{equation}
	I^{(m)}_{\mathbf{r}^+\mathbf{a}_n}=\sum^{d+1}_{ \substack{\abs{\mathbf{b}_n}=1,\\ b_i=\{0,1\}}}C^{(m)}_{\mathbf{r}^+\mathbf{a}_n\to \mathbf{b}_n}I_{\mathbf{b}_n}~~~\label{Basic-Relation-2-2}
\end{equation}
is the one to be determined. Let us focus on the second equation in \eref{Basic-Relation-2} first.
Expanding
\bea
		& & \frac{1}{a_{r}} \frac{\partial}{\partial M_{r}^{2}} I_{\mathbf{a}_{n}}^{(m)} =\frac{1}{a_{r}} \frac{\partial}{\partial M_{r}^{2}}\left[\sum^{d+1}_{ \substack{\abs{\mathbf{b}_n}=1,\\ b_i=\{0,1\}}} C_{\mathbf{a}_{n} \rightarrow \mathbf{b}_{n}}^{(m)} I_{\mathbf{b}_{n}}\right] \nn
		&=&\sum^{d+1}_{ \substack{\abs{\mathbf{b}_n}=1,\\ b_i=\{0,1\}}}\left[\frac{1}{a_{r}}\left( \frac{\partial}{\partial M_{r}^{2}} C_{\mathbf{a}_{n} \rightarrow \mathbf{b}_{n}}^{(m)}\right) I_{\mathbf{b}_{n}}+\frac{1}{a_{r}} C_{\mathbf{a}_{n} \rightarrow \mathbf{b}_{n}}^{(m)} I_{\mathbf{r}^+\mathbf{b}_{n}}\right] \nn
		&=&\sum^{d+1}_{ \substack{\abs{\mathbf{b}_n}=1,\\ b_i=\{0,1\}}}\left[\frac{1}{a_{r}} \left(\frac{\partial}{\partial M_{r}^{2}} C_{\mathbf{a}_{n} \rightarrow \mathbf{b}_{n}}^{(m)} \right)I_{\mathbf{b}_{n}}+\frac{1}{a_{r}} C_{\mathbf{a}_{n} \rightarrow \mathbf{b}_{n}}^{(m)} \sum^{d+1}_{ \substack{\abs{\mathbf{c}_n}=1,\\ c_i=\{0,1\}}} C_{\mathbf{r}^+\mathbf{b}_{n} \rightarrow \mathbf{c}_{n}} I_{\mathbf{c}_{n}}\right]~~~~~~~\label{Basic-Relation-2-3}
\eea
and then
comparing with the \eref{Basic-Relation-2-2}, we have
\begin{equation}
	C^{(m)}_{\mathbf{r}^+\mathbf{a}_n\to \mathbf{b}_n}={1\over a_r}{\partial \over \partial {M_r^2}}C^{(m)}_{\mathbf{a}_n\to \mathbf{b}_n}+{1\over a_r}\sum^{d+1}_{ \substack{\abs{\mathbf{c}_n}=1,\\ c_i=\{0,1\}}}C^{(m)}_{\mathbf{a}_n\to \mathbf{c}_n}C_{\mathbf{r}^+\mathbf{c}_n\to \mathbf{b}_n}\ed~~~~\label{Basic-Relation-2-4}
\end{equation}
The recurrence relation of \eref{Basic-Relation-2-4}
means that to solve  all reductions of $C^{(m)}_{\mathbf{r}^+\mathbf{a}_n\to \mathbf{b}_n}$ one needs to solve just the reduction of  $C_{\mathbf{r}^+\mathbf{c}_n\to \mathbf{b}_n}$ with  $\mathbf{c}_n=\{0,1\},c_r=1$. In other words, we just need to know the reduction
of scalar integrals with just one propagator having the power two. Furthermore,
those scalar reduction coefficients are related to each other by loop momentum shifting $\ell\to \ell+K_r$ and permutation
\begin{equation}
	C_{\mathbf{r}^+\mathbf{b}_n\to \mathbf{c}_n}=\left.C_{\sigma_r(\mathbf{r}^+\mathbf{b}_n)\to \sigma_r(\mathbf{c}_n)}\right\vert_{M_0\leftrightarrow M_r,K_r\to -K_r,K_i\to K_i-K_r}
\end{equation}
where $\sigma_r$ is a permutation defined by $\sigma_r(\mathbf{a}_n)=\{a_r,a_1,\cdots,a_{r-1},a_0,a_{r+1},\cdots,a_n\}$. So we just need to consider the standard scalar reduction coefficient $C_{\underbracket{2,1,1,\cdots,1}_{n+1}\to \mathbf{c}_n}$.

To solve  the reduction of $I_{2,1,1,\cdots,1}$, we use  \eref{Basic-Relation-2} for the particular case
\begin{equation}
	I^{(m)}_{\underbracket{2,1,1,\cdots,1}_{n+1}}=\lim_{D_{n+1}\to D_0}I^{(m)}_{\underbracket{1,1,1,\cdots,1,1}_{n+2}}={\partial \over \partial {M_0^2}}I^{(m)}_{\underbracket{1,1,1,\cdots,1}_{n+1}}\ed~~~~\label{I-2-17}
\end{equation}
%
First using the result in \cite{Feng:2021enk,Hu:2021nia} we can get\footnote{For simplicity, we will write $\sum_{\mathbf{a}_{n+1}=0,1}$
	to represent the sum $\sum^{d+1}_{ \abs{\mathbf{b}_n}=1}\sum_{\mathbf{b}_n=\{0,1\}}$ in \eref{Basic-Relation-2-2}.}
\bea I^{(m)}_{\underbracket{1,1,1,\cdots,1,1}_{n+2}}=\sum_{\mathbf{a}_{n+1}=0,1}C^{(m)}_{\mathbf{1}_{n+2}\to \mathbf{a}_{n+1}}I_{\mathbf{a}_{n+1}}~~~~\label{Basic-Relation-3-1}\eea
where the vector $\mathbf{1}_{n+2}$ means that its all $(n+2)$ components are one.
To take the limit, we write the vector $\mathbf{a}_{n+1}$ as
\bea \mathbf{a}_{n+1}=\{a_0,\mathbf{b}_{n},a_{n+1} \}~~~~\label{frame-3-1}\eea
where its first and $(n+2)$-th component have been explicitly written down and $\mathbf{b}_{n}$ is vector with $n$-components. With this notation we have
\bea & & \lim_{D_{n+1}\to D_0}I^{(m)}_{\underbracket{1,1,1,\cdots,1,1}_{n+2}}=\lim_{D_{n+1}\to D_0}\sum_{a_0, a_{n+1}, \mathbf{b}_{n}=0,1}C^{(m)}_{\mathbf{1}_{n+2}\to \{a_0,\mathbf{b}_{n},a_{n+1} \}}I_{\{a_0,\mathbf{b}_{n},a_{n+1} \}}\nn
& = & \sum_{ \mathbf{b}_{n}=0,1}\WH{C}^{(m)}_{\mathbf{1}_{n+2}\to \{0,\mathbf{b}_{n},0 \}}I_{\{0,\mathbf{b}_{n} \}}
+\sum_{ \mathbf{b}_{n}=0,1}\left(\WH{C}^{(m)}_{\mathbf{1}_{n+2}\to \{1,\mathbf{b}_{n},0 \}}+ \WH{C}^{(m)}_{\mathbf{1}_{n+2}\to \{0,\mathbf{b}_{n},1 \}}\right)I_{\{1,\mathbf{b}_{n} \}}\nn
& & + \sum_{ \mathbf{b}_{n}=0,1}\WH{C}^{(m)}_{\mathbf{1}_{n+2}\to \{1,\mathbf{b}_{n},1 \}}I_{\{2,\mathbf{b}_{n} \}}~~~~\label{frame-3-2}\eea
where $\WH{C}$ means to take the limit of coefficients. It is crucial to notice that
when from the first line to the second line, the limit procedure has been
done for the coefficients and basis separately. This manipulation is legitimate when and only
when the coefficient is not divergent under the limit. Otherwise extra contribution will appear. To avoid this complexity, we should choose the rank $m$ and the limit procedure
properly.

Now we compute the rightest term in \eref{I-2-17}
\bea
& &{\partial \over \partial {M_0^2}}I^{(m)}_{\underbracket{1,1,1,\cdots,1}_{n+1}}={\partial \over \partial {M_0^2}}\sum_{a_{0}=0,1}\sum_{ \mathbf{b}_{n}=0,1}{C}^{(m)}_{\mathbf{1}_{n+1}\to \{a_0,\mathbf{b}_{n} \}}I_{\{a_0,\mathbf{b}_{n} \}}\nn
& = & \sum_{{a}_{0}=0,1}\sum_{ \mathbf{b}_{n}=0,1}\left({\partial \over \partial {M_0^2}}{C}^{(m)}_{\mathbf{1}_{n+1}\to \{a_0,\mathbf{b}_{n} \}}\right)I_{\{a_0,\mathbf{b}_{n} \}}
+ \sum_{ \mathbf{b}_{n}=0,1}{C}^{(m)}_{\mathbf{1}_{n+1}\to \{1,\mathbf{b}_{n} \}}I_{\{2,\mathbf{b}_{n} \}}\nn
& = & \sum_{ \mathbf{b}_{n}=0,1}\left({\partial \over \partial {M_0^2}}{C}^{(m)}_{\mathbf{1}_{n+1}\to \{0,\mathbf{b}_{n} \}}\right)I_{\{0,\mathbf{b}_{n} \}}
+ \sum_{ \mathbf{b}_{n}=0,1}\left({\partial \over \partial {M_0^2}}{C}^{(m)}_{\mathbf{1}_{n+1}\to \{1,\mathbf{b}_{n} \}}\right)I_{\{1,\mathbf{b}_{n} \}}\nn
&  & +\sum_{ \mathbf{b}_{n}=0,1}{C}^{(m)}_{\mathbf{1}_{n+1}\to \{1,\mathbf{b}_{n} \}}I_{\{2,\mathbf{b}_{n} \}}\ed~~~~\label{frame-3-3}
\eea
Combining \eref{frame-3-2} and \eref{frame-3-3}, we get
\bea & & \left({C}^{(m)}_{\mathbf{1}_{n+1}\to \{1,\mathbf{1}_{n} \}}-\WH{C}^{(m)}_{\mathbf{1}_{n+2}\to \{1,\mathbf{1}_{n},1 \}}\right)I_{\{2,\mathbf{1}_{n} \}}\nn
& = & \sum_{ \mathbf{b}_{n}=0,1}\left(\WH{C}^{(m)}_{\mathbf{1}_{n+2}\to \{0,\mathbf{b}_{n},0 \}}-{\partial \over \partial {M_0^2}}{C}^{(m)}_{\mathbf{1}_{n+1}\to \{0,\mathbf{b}_{n} \}}\right)I_{\{0,\mathbf{b}_{n} \}}\nn
& &
+\sum_{ \mathbf{b}_{n}=0,1}\left(\WH{C}^{(m)}_{\mathbf{1}_{n+2}\to \{1,\mathbf{b}_{n},0 \}}+ \WH{C}^{(m)}_{\mathbf{1}_{n+2}\to \{0,\mathbf{b}_{n},1 \}}-{\partial \over \partial {M_0^2}}{C}^{(m)}_{\mathbf{1}_{n+1}\to \{1,\mathbf{b}_{n} \}}\right)I_{\{1,\mathbf{b}_{n} \}}\nn
& & + \sum^\prime_{ \mathbf{b}_{n}=0,1}\left(\WH{C}^{(m)}_{\mathbf{1}_{n+2}\to \{1,\mathbf{b}_{n},1 \}}-{C}^{(m)}_{\mathbf{1}_{n+1}\to \{1,\mathbf{b}_{n} \}}\right)I_{\{2,\mathbf{b}_{n} \}}~~~~\label{frame-3-4}
\eea
where the prime summation of the last line in \eref{frame-3-4} is to denote that we should exclude the case that $\mathbf{b}_{n}=\mathbf{1}_{n}$. Noticing that also because the last line in \eref{frame-3-4}, the reduction of scalar integrals with the first propagator having power two
is in the recurrence pattern by comparing $I_{\{2,\mathbf{1}_{n} \}}$ and $I_{\{2,\mathbf{b}_{n} \}}$. Since we know the reduction of lower topologies, the left hand side
of \eref{frame-3-4} can be solved if its coefficient $\left({C}^{(m)}_{\mathbf{1}_{n+1}\to \{1,\mathbf{1}_{n} \}}-\WH{C}^{(m)}_{\mathbf{1}_{n+2}\to \{1,\mathbf{1}_{n},1 \}}\right)$ is not zero. To guarantee this point, the rank $m$ and the limit procedure should be chosen properly. 


\subsection{{Choice of $R$}}

Having set up the general frame, we could use it to compute the reduction of other topologies, i.e., the bubble, triangle, box and pentagon. However, before doing so, let us clarify some points regarding to \eref{frame-3-4}. First in \eref{frame-3-4}, there is a tensor rank $m$ we need to choose.
Naively, starting from $(n+2)$ propagators, we should choose  $m\geq n+1$ since only with this choice, the reduction will go down to tadpoles. 
However, we know the final result should be independent with the choice of $m$, thus $m< n+1$ should be fine. 
The reason that we could get the tadpole reduction coefficients comes from the last term at the RHS of \eref{frame-3-4} where recurrence between different $n$ with {propagators having general powers} is manifest. Although the rank $m$ can be chosen arbitrarily, the coefficient at the LHS could be zero for a particular choice of $m$.
For  example, for  the tadpole discussed before, if we take $m=1$, one can find
$C^{(1)}_{1\to 1}=0$ and $C^{(1)}_{1,1\to 1,1}={(M_0^2-M_1^2+K_1^2) R\cdot K_1\over K_1^2}$.
Under the limit, $M_1\to M_0, K_1\to 0$, we have $C^{(1)}_{1,1\to 1,1}\to R\cdot K_1\to 0$, thus we have coefficient $\left({C}^{(m)}_{\mathbf{1}_{n+1}\to \{1,\mathbf{1}_{n} \}}-\WH{C}^{(m)}_{\mathbf{1}_{n+2}\to \{1,\mathbf{1}_{n},1 \}}\right)=0$ at the left hand side of \eref{frame-3-4}.

Except the independence of rank $m$, the reduction is also independent of the auxiliary vector $R$. This observation gives us the freedom to choose $R$ to simplify the whole
computation greatly. A  particular choice of the limit procedure, which will avoid the singularity of \eref{frame-3-4}, is following:
\begin{equation}
	\begin{cases}
		R \cdot K_{i}=0, \forall i \leq n,  \\
		\lim _{D_{n+1} \rightarrow D_{0}}=\lim _{M_{n+1} \rightarrow M_{0}, K_{n+1} \rightarrow t R, t \rightarrow 0}\ed
	\end{cases}~~~\label{gauge-1}
\end{equation}
For these two conditions, the first one is the special choice of $R$ while the second one tells us how to take the limit.
With this choice, one can find:
\begin{equation}
	\begin{gathered}
		\lim _{D_{n+1} \rightarrow D_{0}}\left( f_{i}=K_{i} \cdot K_{i}+M_{0}^{2}-M_{i}^{2}\right)= \begin{cases}f_{i} & \text { if } i\leq n \\
			s_{(n+1)(n+1) } & \text { if } i=n+1\end{cases} \\
		\lim _{D_{n+1} \rightarrow D_{0}} \left(s_{0 i}=K_{i} \cdot R \right) = \begin{cases}0 & \text { if } 0<i\leq n \\
			\lim _{t \rightarrow 0} t s_{00} & \text { if } i=n+1\ed\end{cases}
	\end{gathered}~~~~\label{gauge-2}
\end{equation}

Now we discuss the choice of rank $m$. First, $m=1$ will lead zero coefficient at the LHS of  \eref{frame-3-4} as shown in the tadpole example. To simplify our calculation, we require that $R$ appears as few as possible. With the  choice \eref{gauge-1}, the only nonzero contraction of $R$ is $R^2=s_{00}$, so it is natural to choose $m=2$, which guarantees $s_{00}$ only appearing once in every term. For  the rank $m=2$,  by our general discussion, the allowed Lorentz invariant combinations before the limit are only $s_{0i}=K_i\cdot R$ and $s_{00}=R^2$ and all reduction coefficients are the form
\bea A R^2+\sum_{ij} B_{ij} (K_i\cdot R)(K_j\cdot R)\ed~~~\label{gauge-3}\eea
With the choice \eref{gauge-1}, we have
\bea A R^2+\sum_{ij} B_{ij} (K_i\cdot R)(K_j\cdot R)
\to \left\{ \begin{array}{ll} AR^2,~~~ & {\rm for}~ {C}^{(2)}_{\mathbf{1}_{n+1}\to \{\mathbf{b}_{n+1} \}}\\
	A R^2+ B_{(n+1)(n+1)} t^2(R\cdot R)^2,~~~ & {\rm for}~{C}^{(2)}_{\mathbf{1}_{n+2}\to \{\mathbf{b}_{n+2} \}}\end{array}\right.~~~~~~\label{gauge-4}\eea
The $A$ and $B$ are two coefficients, which have been solved in   \cite{Feng:2021enk,Hu:2021nia}. The general expression for $A$ is
\begin{equation}
	A=M_{0}^{2} \frac{1}{D-n}\left(4-\alpha_{n}^{T} \widetilde{\boldsymbol{G}}^{-1} \alpha_{n}\right)~~~\label{A-gen}
\end{equation}
where $\alpha_{n}$ is a vector defined as
\begin{equation}
	\alpha_{n}^{T}=\left(\alpha_{1}, \alpha_{2}, \cdots, \alpha_{n}\right)=\left(\frac{f_{1}}{M_{0}^{2}}, \frac{f_{2}}{M_{0}^{2}}, \cdots, \frac{f_{n}}{M_{0}^{2}}\right)\co
\end{equation}
and $\widetilde{G}=\left[s_{i j} / M_{0}^{2}\right]$ is the $n \times n$ rescaled Gram matrix.
Thus the first line of \eref{gauge-4} is known. Now we consider the second line.
By taking $R=K_{n+1}/t$, we have
\bea
\left.I^{(2)}_{\underbracket{1,1,1,\cdots,1}_{n+2}}\right\vert_{R=K_{n+1}/t}=f_{n+1}^2t^{-2}I_{\underbracket{1,1,1,\cdots,1}_{n+2}}+\text{Lower\ topologies}
\eea
where the $f_{n+1}^2t^{-2}$ is the reduction coefficient under the choice \eref{gauge-1}. Thus we find
\begin{align}
	\widehat{C}_{\mathbf{1_{n+2}} \rightarrow\left\{1,\mathbf{1_{n}}, 1\right\}}^{(2)}=\lim_{t\to 0,f_{n+1}=t^2 s_{00}}f_{n+1}^2t^{-2}=\lim_{t\to 0}t^2 s_{00}=0\ed
\end{align}
Putting all together we have
\bea \left(C_{\mathbf{1_{n+1}} \rightarrow\left\{1,\mathbf{1_{n}}\right\}}^{(2)}-\widehat{C}_{\mathbf{1_{n+2}} \rightarrow\left\{1,\mathbf{1_{n}}, 1\right\}}^{(2)}\right)= C_{\mathbf{1_{n+1}} \rightarrow\left\{1,\mathbf{1_{n}}\right\}}^{(2)}= A R^2\ed\eea
Substituting the result to \eref{frame-3-4} we get
\bea  I_{\{2,\mathbf{1}_{n} \}}&=&{1\over A R^2}\left\{
\sum_{ \mathbf{b}_{n}=0,1}\left(\WH{C}^{(2)}_{\mathbf{1}_{n+2}\to \{0,\mathbf{b}_{n},0 \}}-{\partial \over \partial {M_0^2}}{C}^{(2)}_{\mathbf{1}_{n+1}\to \{0,\mathbf{b}_{n} \}}\right)I_{\{0,\mathbf{b}_{n} \}}\right.\nn
& &
+\sum_{ \mathbf{b}_{n}=0,1}\left(\WH{C}^{(2)}_{\mathbf{1}_{n+2}\to \{1,\mathbf{b}_{n},0 \}}+ \WH{C}^{(2)}_{\mathbf{1}_{n+2}\to \{0,\mathbf{b}_{n},1 \}}-{\partial \over \partial {M_0^2}}{C}^{(2)}_{\mathbf{1}_{n+1}\to \{1,\mathbf{b}_{n} \}}\right)I_{\{1,\mathbf{b}_{n} \}}\nn
& & \left. + \sum^\prime_{ \mathbf{b}_{n}=0,1}\left(\WH{C}^{(2)}_{\mathbf{1}_{n+2}\to \{1,\mathbf{b}_{n},1 \}}-{C}^{(2)}_{\mathbf{1}_{n+1}\to \{1,\mathbf{b}_{n} \}}\right)I_{\{2,\mathbf{b}_{n} \}}\right\}\co ~~~~\label{frame-3-5}
\eea
Where at the RHS the  choice \eref{gauge-1} is assumed. It will be apparent from later examples that although there is ${1\over R^2}$ in the front of \eref{frame-3-5}, all coefficients will contain the $R^2$ factor under the limit choice \eref{gauge-2}, so the $R$-dependent will be cancelled eventually. It will become a consistent check of our method.

\section{Examples}

Having set up the general frame in \eref{frame-3-5}, we compute the reduction of scalar integrals with just one propagator having power two.
Since the tadpole has been done in \eref{tad-8}, we will focus on  other topologies, i.e., the bubble, triangle, box and pentagon. In this section, we will give the details for bubble and triangle  for further demonstration of  the method.
The results for box and pentagon will be presented in the Appendix and the attached Mathematica file. In this section, we will use the known reduction coefficients with tensor rank $2$ for various topologies, which can be found in \cite{Feng:2021enk,Hu:2021nia}.

\subsection{The bubble}

Now we consider the reduction of bubble $I_{2,1}$.
First we have (see \eref{A-gen})
\bea {1\over A R^2} & = & \frac{1-D}{s_{00} M_{0}^{2}\left(4-\alpha_{1}^{T} \widetilde{\boldsymbol{G}}^{-1} \alpha_{1}\right)}= { (1-D)\over -s_{00} \frac{-2 M_{0}^{2}\left(s_{11}+M_{1}^{2}\right)+\left(M_{1}^{2}-s_{11}\right)^{2}+M_{0}^{4}}{s_{11}}}\eea
where $s_{00}=R^2$ and $s_{ij}=K_i\cdot K_j$.
Now we compute various coefficients of $I$ in \eref{frame-3-5}:
\begin{itemize}

    \item {\bf $I_{0, 1}$}: The coresponding coefficients are
    \bea
    & &\widehat{C}^{(2)}_{\{1,1,1\}\to \{0,1,0\}}\to -\frac{s_{00}}{s_{11}},~~~~ \partial_{M_{0}^{2}}C^{(2)}_{\{1,1\}\to \{0,1\}}\to  -\frac{s_{00}}{(D-1) s_{11}}
    \eea
    under the  choice \eref{gauge-1}, i.e., $s_{0i}=0,i<2$.

    \item {\bf $I_{1, 0}$}: The coresponding coefficients are
    \bea
    & &\widehat{C}^{(2)}_{\{1,1,1\}\to \{0,0,1\}}\to  \frac{s_{00}}{s_{11}},~~~~~
    \widehat{C}^{(2)}_{\{1,1,1\}\to \{1,0,0\}}\to  0,~~~~ \partial_{M_{0}^{2}}C^{(2)}_{\{1,1\}\to \{1,0\}}\to  \frac{s_{00}}{(D-1) s_{11}}
    \ed\eea

    \item {\bf $I_{1, 1}$}: The coresponding coefficients are
    \bea
    & &\widehat{C}^{(2)}_{\{1,1,1\}\to \{0,1,1\}}\to  \frac{s_{00} \left(2 s_{11}-M_0^2+M_1^2\right)}{s_{11}},~~~~
    \widehat{C}^{(2)}_{\{1,1,1\}\to \{1,1,0\}}\to  -s_{00},~~~\nn%
    & & \partial_{M_{0}^{2}}C^{(2)}_{\{1,1\}\to \{1,1\}}\to \frac{2 s_{00} \left(s_{11}-M_0^2+M_1^2\right)}{(D-1) s_{11}}
    \ed\eea

    \item {\bf $I_{2, 0}$}: The coresponding coefficients are
    \bea
    & &\widehat{C}^{(2)}_{\{1,1,1\}\to \{1,0,1\}}\to 0,~~~
   {C}^{(2)}_{\{1,1\}\to \{1,0\}}\to  \frac{s_{00} \left(s_{11}+M_0^2-M_1^2\right)}{(D-1) s_{11}}
   \ed\eea

\end{itemize}
When combining all together and using the result \eref{tad-8}  we get

\bea I_{2,1} =
\sum^2_{ \abs{\mathbf{b}_n}=1}\sum_{\mathbf{b}_n=\{0,1\}}C_{{2,1}\to \mathbf{b}_n}I_{\mathbf{b}_n}\eea
with
\bea C_{{2,1}\to {1,1}} & = & {2 (D-3) M_0^2 \left(M_0^2-M_1^2-s_{11}\right)\over \a_{2,1}} \nn
C_{{2,1}\to {1,0}} & = & {-(D-2) \left(M_0^2+M_1^2-s_{11}\right)\over \a_{2,1}} \nn
	C_{{2,1}\to {1,1}} & = & {2 (D-2) M_0^2\over \a_{2,1}} \nn
	\a_{2,1}&= & {2 M_0^2 \left(M_0^2-2 M_1 M_0+M_1^2-s_{11}\right) \left(M_0^2+2 M_1 M_0+M_1^2-s_{11}\right)}\ed \eea
%

\subsection{The triangle}

For the triangle, we have
\bea {1\over AR^2}& = & \frac{2-D}{s_{00} M_{0}^{2}\left(4-\alpha_{2}^{T} \widetilde{\boldsymbol{G}}^{-1} \alpha_{2}\right)}= {(D-2) \left(s_{11} s_{22}-s_{12}^2\right)\over s_{00}\a_{2,1,1}}\eea
with
\bea & & \a_{2,1,1}=M_0^4 \left(s_{11}-2 s_{12}+s_{22}\right)-2 M_0^2 \left(s_{12} \left(s_{11}-2 s_{12}+s_{22}\right)+M_2^2 \left(s_{11}-s_{12}\right)+M_1^2 \left(s_{22}-s_{12}\right)\right)\nn%
& & -2 M_2^2 \left(s_{11} \left(s_{22}-s_{12}\right)+M_1^2 s_{12}\right)+s_{22} \left(s_{11} \left(s_{11}-2 s_{12}+s_{22}\right)-2 M_1^2 \left(s_{11}-s_{12}\right)+M_1^4\right)+M_2^4 s_{11}~~~~~~\ed\eea
Other coefficients are given by:
\begin{itemize}
  \item {\bf $I_{0, 0, 1}$}: The coresponding coefficients are
  \bea
  & &\widehat{C}^{(2)}_{\{1,1,1,1\}\to \{0,0,1,0\}}\to  0,~~~
 \partial_{M_0^2}C^{(2)}_{\{1,1,1\}\to \{0,0,1\}}\to  0
 \ed\eea
      under the  choice \eref{gauge-1}, i.e., $s_{0i}=0,i<3$.
  \item {\bf $I_{0, 1, 0}$}: The coresponding coefficients are
  \bea
  & &\widehat{C}^{(2)}_{\{1,1,1,1\}\to \{0,1,0,0\}}\to  0,~~~
  \partial_{M_0^2}C^{(2)}_{\{1,1,1\}\to \{0,1,0\}}\to  0
  \ed\eea

  \item {\bf $I_{0, 1, 1}$}: The coresponding coefficients are
  \bea
  \widehat{C}^{(2)}_{\{1,1,1,1\}\to \{0,1,1,0\}}\to -\frac{s_{00} \left(s_{11}-2 s_{12}+s_{22}\right)}{s_{11} s_{22}-s_{12}^2},~~
  \partial_{M_0^2}C^{(2)}_{\{1,1,1\}\to \{0,1,1\}}\to\frac{s_{00} \left(s_{11}-2 s_{12}+s_{22}\right)}{(D-2) \left(s_{12}^2-s_{11} s_{22}\right)}~~~~~~~
  \ed\eea

  \item {\bf $I_{1, 0, 0}$}: The coresponding coefficients are
  \bea
  \widehat{C}^{(2)}_{\{1,1,1,1\}\to \{0,0,0,1\}}\to 0,~
  \widehat{C}^{(2)}_{\{1,1,1,1\}\to \{1,0,0,0\}}\to 0,~
  \partial_{M_0^2}C^{(2)}_{\{1,1,1\}\to \{1,0,0\}}\to 0
  \ed\eea

  \item {\bf $I_{1, 0, 1}$}: The coresponding coefficients are
  \bea
  & &\widehat{C}^{(2)}_{\{1,1,1,1\}\to \{0,0,1,1\}}\to \frac{s_{00} \left(s_{12}-s_{22}\right)}{s_{12}^2-s_{11} s_{22}},~~~~
  \widehat{C}^{(2)}_{\{1,1,1,1\}\to \{1,0,1,0\}}\to 0\nn
  & &\partial_{M_0^2}C^{(2)}_{\{1,1,1\}\to \{1,0,1\}}\to \frac{s_{00} \left(s_{12}-s_{22}\right)}{(D-2) \left(s_{12}^2-s_{11} s_{22}\right)}
  \ed\eea

  \item {\bf $I_{1, 1, 0}$}: The coresponding coefficients are
  \bea
  & &\widehat{C}^{(2)}_{\{1,1,1,1\}\to \{0,1,0,1\}}\to \frac{s_{00} \left(s_{11}-s_{12}\right)}{s_{11} s_{22}-s_{12}^2},~~~
  \widehat{C}^{(2)}_{\{1,1,1,1\}\to \{1,1,0,0\}}\to 0\nn
  & &\partial_{M_0^2}C^{(2)}_{\{1,1,1\}\to \{1,1,0\}}\to \frac{s_{00} \left(s_{12}-s_{11}\right)}{(D-2) \left(s_{12}^2-s_{11} s_{22}\right)}
  \ed\eea

  \item {\bf $I_{1, 1, 1}$}: For this case, the corresponding coefficients are little bit long.
  For the simplicity of the presentation, we will write down $C^{N}$ and  $C^{D}$ separately and 
  understand $C={C^{N}\over C^{D}}$:
  \bea  
  \widehat{C}^{(2),N}_{\{1,1,1,1\}\to  \{0,1,1,1\}} &\to & -s_{00} \left(s_{11} \left(s_{12}+s_{22}-M_0^2+M_2^2\right)+s_{12} \left(s_{22}+2 M_0^2-M_1^2-M_2^2\right)\right.
  	 \nn 
  & & \left.-3 s_{12}^2+\left(M_1^2-M_0^2\right) s_{22}\right)\nn
  \widehat{C}^{(2),D}_{\{1,1,1,1\} \to   \{0,1,1,1\}}&\to & {s_{12}^2-s_{11} s_{22}}\eea
  and
  \bea \widehat{C}^{(2)}_{\{1,1,1,1\}\to \{1,1,1,0\}}\to -s_{00}\eea
  and 
  \bea \partial_{M_0^2}C^{(2),N}_{\{1,1,1\}\to \{1,1,1\}}& \to & 2 s_{00} \left(-s_{11} \left(s_{12}-M_0^2+M_2^2\right)+s_{12} \left(-s_{22}-2 M_0^2+M_1^2+M_2^2\right)\right. \nn & & \left. ~~~~+2 s_{12}^2+\left(M_0^2-M_1^2\right) s_{22}\right)\nn
  \partial_{M_0^2}C^{(2),D}_{\{1,1,1\}\to \{1,1,1\}}& \to & {(D-2) \left(s_{12}^2-s_{11} s_{22}\right)}\ed\eea

  \item {\bf $I_{2, 0, 0}$}: The coresponding coefficients are
  \bea
  & &\widehat{C}^{(2)}_{\{1,1,1,1\}\to \{1,0,0,1\}}\to  0,~~~~
  {C}^{(2)}_{\{1,1,1\}\to \{1,0,0\}}\to  0
  \ed\eea

  \item {\bf $I_{2, 0, 1}$}: The coresponding coefficients are
  \bea
  & &\widehat{C}^{(2)}_{\{1,1,1,1\}\to \{1,0,1,1\}}\to 0,\nn
  & & {C}^{(2)}_{\{1,1,1\}\to \{1,0,1\}}\to  \frac{s_{00} \left(-s_{11} s_{22}+s_{12} \left(s_{22}+M_0^2-M_2^2\right)+\left(M_1^2-M_0^2\right) s_{22}\right)}{(D-2) \left(s_{12}^2-s_{11} s_{22}\right)}~~~~
  \ed\eea

  \item {\bf $I_{2, 1, 0}$}: The coresponding coefficients are
  \bea
  & &\widehat{C}^{(2)}_{\{1,1,1,1\}\to \{1,1,0,1\}}\to 0\\
  & &{C}^{(2)}_{\{1,1,1\}\to \{1,1,0\}}\to  \frac{s_{00} \left(s_{11} \left(s_{12}-s_{22}-M_0^2+M_2^2\right)+\left(M_0^2-M_1^2\right) s_{12}\right)}{(D-2) \left(s_{12}^2-s_{11} s_{22}\right)}
  \ed\eea
\end{itemize}

When combining all together and simplify we get
\bea I_{2,1,1} =
\sum^3_{ \abs{\mathbf{b}_n}=1}\sum_{\mathbf{b}_n=\{0,1\}}C_{{2,1  ,1}\to \mathbf{b}_n}I_{\mathbf{b}_n}\ed\eea
with
\bea   
  && C_{{2,1,1}\to{0, 0, 1}}  =  \frac{(D-2) \left(s_{22} \left(s_{11}-s_{12}+M_0^2-M_1^2\right)+\left(M_2^2-M_0^2\right) s_{12}\right)}{\a_{2,1,1} \left(-2 M_0^2 \left(s_{22}+M_2^2\right)+\left(M_2^2-s_{22}\right){}^2+M_0^4\right)}\nn
  && C_{{2,1,1}\to{0, 1, 0}}  =  \frac{(D-2) \left(-s_{11} \left(s_{12}-s_{22}+M_2^2\right)+M_0^2 \left(s_{11}-s_{12}\right)+M_1^2 s_{12}\right)}{\a_{2,1,1} \left(-2 M_0^2 \left(s_{11}+M_1^2\right)+\left(M_1^2-s_{11}\right){}^2+M_0^4\right)}\nn
  && C_{{2,1,1}\to{0, 1, 1}}  =  \frac{(D-3) \left(s_{11}-2 s_{12}+s_{22}\right)}{\a_{2,1,1}}\nn
\eea
For the long corresponding coefficients, we write them as: 
\bea
  && C_{{2,1,1}\to{1, 0, 0}} = \frac{C_{{2,1,1}\to{1, 0, 0}}^{(2),N} }{C_{{2,1,1}\to{1, 0, 0}}^{(2),D} } \\
  && \longline{C_{{2,1,1}\to{1, 0, 0}}^{(2),N}}{-(D-2) (M_0^8 (s_{11}-2 s_{12}+s_{22})+M_0^6 (M_1^2 (s_{11}+2 s_{12}-3 s_{22})+M_2^2 (-3 s_{11}+2 s_{12}+s_{22})-(s_{11}+s_{22}) (s_{11}-2 s_{12}+s_{22}))+M_0^2 (M_1^4 (s_{22} (3 s_{11}-2 s_{12}-3 s_{22})+M_2^2 (3 s_{22}-2 s_{12}))+M_1^2 (4 M_2^2 (s_{11} (s_{12}-s_{22})+s_{12} s_{22})+s_{22} (s_{11} (4 s_{12}+s_{22})-3 s_{11}^2-2 s_{12} s_{22}) +M_2^4 (3 s_{11}-2 s_{12}))+s_{11} (M_2^4 (-3 s_{11}-2 s_{12}+3 s_{22})+M_2^2 (s_{11} (s_{22}-2 s_{12})+s_{22} (4 s_{12}-3 s_{22}))+s_{22} (s_{11}+s_{22}) (s_{11}-2 s_{12}+s_{22})-M_2^6)+M_1^6 (-s_{22}))-(M_1^2-s_{11}) (M_2^2-s_{22}) (-2 M_2^2 (s_{11} (s_{22}-s_{12})+M_1^2 s_{12})+s_{22} ((M_1^2-s_{11}) (-s_{11}+2 s_{12}+M_1^2)+s_{11} s_{22})+M_2^4 s_{11})+3 M_0^4 (-s_{11}+s_{22}+M_1^2-M_2^2) (M_1^2 s_{22}-M_2^2 s_{11}))}\no \\
  && \longline{C_{{2,1,1}\to{1, 0, 0}}^{(2),D} }{ 2 \a_{2,1,1} M_0^2 ((M_0-M_1){}^2-s_{11}) ((M_0+M_1){}^2-s_{11}) ((M_0-M_2){}^2-s_{22}) ((M_0+M_2){}^2-s_{22})}\no
\eea
\bea
  && C_{{2,1,1}\to{1, 0, 1}} = \frac{C_{{2,1,1}\to{1, 0, 1}}^{(2),N} }{C_{{2,1,1}\to{1, 0, 1}}^{(2),D} } \\
  && \longline{C_{{2,1,1}\to{1, 0, 1}}^{(2),N}}{(D-3) s_{22} (M_0^2 (s_{11}-2 s_{12}+s_{22}-M_1^2+M_2^2)+s_{22} (-s_{11}+2 s_{12}+M_1^2+2 M_2^2)-M_2^2 (s_{11}+2 s_{12}-M_1^2+M_2^2)-s_{22}^2)}\no \\
  && \longline{C_{{2,1,1}\to{1, 0, 1}}^{(2),D}}{\a_{2,1,1} (-2 M_0^2 (s_{22}+M_2^2)+(M_2^2-s_{22}){}^2+M_0^4)}\no
\eea
\bea
  && C_{{2,1,1}\to{1, 1, 0}} = \frac{C_{{2,1,1}\to{1, 1, 0}}^{(2),N} }{C_{{2,1,1}\to{1, 1, 0}}^{(2),D} } \\
  && \longline{C_{{2,1,1}\to{1, 1, 0}}^{(2),N}}{(D-3) s_{11} (M_1^2 (-2 s_{11}+2 s_{12}+s_{22}-M_2^2)+s_{11} (s_{11}-2 s_{12}+s_{22}-M_2^2)-M_0^2 (s_{11}-2 s_{12}+s_{22}+M_1^2-M_2^2)+M_1^4)}\no \nn
  && \longline{C_{{2,1,1}\to{1, 1, 0}}^{(2),D}}{\a_{2,1,1} (-2 M_0^2 (s_{11}+M_1^2)+(M_1^2-s_{11}){}^2+M_0^4)}\no
\eea
\bea
  && C_{{2,1,1}\to{1, 1, 1}} = \frac{C_{{2,1,1}\to{1, 1, 1}}^{(2),N} }{\a_{2,1,1}} \\
  && \longline{C_{{2,1,1}\to{1, 1, 1}}^{(2),N}}{(D-4) (M_0^2 (s_{11}-2 s_{12}+s_{22})+M_2^2 (s_{12}-s_{11})+s_{12} (-s_{11}+2 s_{12}+M_1^2)-s_{22} (s_{12}+M_1^2))}\no
\eea

\section{Conclusion}

In this paper, we have shown how to use the PV-reduction method with auxiliary vector $R$ for propagators with power one in \cite{Feng:2021enk,Hu:2021nia} to get the general tensor reduction, i.e., with arbitrary tensor structure in the numerator and general power for propagators in the denominator.

Our idea is to consider tensor reduction of the topology with just one more propagator and then take proper limits. We have used several examples to demonstrate our method. The examples of bubble and triangle have been given in the main text, while the partial results for the box and pentagon have been collected in the Appendix. Furthermore, we have provided a Mathematica file containing all results for box and pentagon. All results have been checked with \textbf{FIRE6} \cite{Smirnov:2008iw,Smirnov:2014hma,Smirnov:2019qkx}.
Indeed, the IBP algorithm can be used to do reduction for general one-loop integrals having propagators with the general power as programmed by FIRE and Kira \cite{Maierhofer:2017gsa,Klappert:2020nbg,Lange:2021edb}. However, one of the key results emphasized in this paper is that we do not need to use the IBP algorithm and our improved PV-reduction method can do the same thing, thus it is a self-complete reduction method for general one-loop integrals.
Another good point of our method is that with the established recurrence relations for reduction coefficients of different tensor ranks, one can obtain reduction coefficients for very higher tensor ranks with much less efforts, which will become very time consuming if one tries to solve IBP relations directly.   

With the result in this paper, we have demonstrated that our improved PV-reduction method, i.e., PV-reduction method with auxiliary vector, is a self-completed reduction method for the one-loop integrals, just like the traditional IBP method. It is obvious that the next step is to generalize the idea to higher loops. If we know the reduction for propagators with power one for higher loops, we can get the reduction for propagators with general powers using the same idea.

\section*{Acknowledgments}

This work is supported by Chinese NSF funding under Grant No.11935013, No.11947301, No.12047502 (Peng Huanwu Center).

\appendix
\section{The box}

Here we exhibit partial results for the box and pentagon. We only give the intermediate coefficients such as \eqref{frame-3-5}. The final reduction coefficients are too long, so we collect the complete results in a Mathematica file and upload it in \href{https://github.com/Wanghongbin123/HigherPole}{Github} for easy reference.

First we have 
\bea
 \frac{1}{AR^2}=\frac{(3-D) \left(s_{11} \left(s_{23}^2-s_{22} s_{33}\right)-2 s_{12} s_{13} s_{23}+s_{12}^2 s_{33}+s_{13}^2 s_{22}\right)}{s_{00} \a_{2,1,1,1}}
\eea
where
\bea
\longline{\a_{2,1,1,1}}{(4 M_0^2 (s_{11} (s_{23}^2-s_{22} s_{33})-2 s_{12} s_{13} s_{23}+s_{12}^2 s_{33}+s_{13}^2 s_{22})+(s_{33}+M_0^2-M_3^2) ((s_{22}+M_0^2-M_2^2) (s_{12} s_{13}-s_{11} s_{23})-(s_{11}+M_0^2-M_1^2) (s_{13} s_{22}-s_{12} s_{23})-(s_{33}+M_0^2-M_3^2) (s_{12}^2-s_{11} s_{22}))+(s_{22}+M_0^2-M_2^2) ((s_{33}+M_0^2-M_3^2) (s_{12} s_{13}-s_{11} s_{23})+(s_{11}+M_0^2-M_1^2) (s_{13} s_{23}-s_{12} s_{33})-(s_{22}+M_0^2-M_2^2) (s_{13}^2-s_{11} s_{33}))-(s_{11}+M_0^2-M_1^2) ((s_{11}+M_0^2-M_1^2) (s_{23}^2-s_{22} s_{33})+(s_{33}+M_0^2-M_3^2) (s_{13} s_{22}-s_{12} s_{23})-(s_{22}+M_0^2-M_2^2) (s_{13} s_{23}-s_{12} s_{33})))}
\ed\eea
For other terms:
\begin{itemize} 
  \item {\bf $I_{0, 0, 0, 1}$}: The coresponding coefficients are
  \bea 
  &&\widehat{C}_{\{1,1,1,1,1\}\to \{0,0,0,1,0\}}\to 0 \nn
  &&\partial_{M_0^2}C_{\{1,1,1,1\}\to \{0,0,0,1\}}\to 0
  \ed\eea 
  There is only one permutation transformation between $I_{0, 0, 1, 0}$,$I_{0, 1, 0, 0}$,$I_{1, 0, 0, 0}$ and $I_{0, 0, 0, 1}$, so they are all 0.

  \item {\bf $I_{0, 0, 1, 1}$}: The coresponding coefficients are
  \bea 
  &&\widehat{C}_{\{1,1,1,1,1\}\to \{0,0,1,1,0\}}\to 0  \nn
  &&\partial_{M_0^2}C_{\{1,1,1,1\}\to \{0,0,1,1\}}\to 0
  \ed\eea   
  According to permutation symmetry, $I_{0, 1, 0, 1}$, $I_{0, 1, 1, 0}$, $I_{1, 0, 0, 1}$, $I_{1, 0, 1, 0}$ and $I_{1, 1, 0, 0}$ are all 0. 
  These coefficients vanish because the starting pentagon will not be reduced to tadpoles and bubbles for tensor rank $2$.
  
  \item {\bf $I_{0, 1, 1, 1}$}: The coresponding coefficients are
  \bea 
  &&\widehat{C}_{\{1,1,1,1,1\}\to \{0,1,1,1,0\}}\to\frac{\widehat{C}_{\{1,1,1,1,1\}\to \{0,1,1,1,0\}}^{(2),N}}{\widehat{C}_{\{1,1,1,1,1\}\to \{0,1,1,1,0\}}^{d+1}}\nn
  &&\longline{\widehat{C}_{\{1,1,1,1,1\}\to \{0,1,1,1,0\}}^{(2),N}}{-s_{00} (-s_{11} s_{22}+2 s_{11} s_{23}-s_{11} s_{33}-2 s_{12} (s_{13}+s_{23}-s_{33})+s_{12}^2+2 s_{13} (s_{22}-s_{23})+s_{13}^2-s_{22} s_{33}+s_{23}^2)}\nn
  &&\longline{\widehat{C}_{\{1,1,1,1,1\}\to \{0,1,1,1,0\}}^{(2),D}}{s_{11} (s_{23}^2-s_{22} s_{33})-2 s_{12} s_{13} s_{23}+s_{12}^2 s_{33}+s_{13}^2 s_{22}}\\
  \nn
  &&\partial_{M_0^2}C_{\{1,1,1,1\}\to \{0,1,1,1\}}\to\frac{\partial_{M_0^2}C_{\{1,1,1,1\}\to \{0,1,1,1\}}^{(2),N}}{\partial_{M_0^2}C_{\{1,1,1,1\}\to \{0,1,1,1\}}^{(2),D}}\nn
  &&\longline{\partial_{M_0^2}C_{\{1,1,1,1\}\to \{0,1,1,1\}}^{(2),N}}{-s_{00} (-s_{11} s_{22}+2 s_{11} s_{23}-s_{11} s_{33}-2 s_{12} (s_{13}+s_{23}-s_{33})+s_{12}^2+2 s_{13} (s_{22}-s_{23})+s_{13}^2-s_{22} s_{33}+s_{23}^2)}\nn
  &&\longline{\partial_{M_0^2}C_{\{1,1,1,1\}\to \{0,1,1,1\}}^{(2),D}}{(D-3) (s_{11} (s_{23}^2-s_{22} s_{33})-2 s_{12} s_{13} s_{23}+s_{12}^2 s_{33}+s_{13}^2 s_{22})}\nonumber
  \eea    
    
  \item {\bf $I_{1, 0, 1, 1}$}: The coresponding coefficients are
  \bea 
  &&\widehat{C}_{\{1,1,1,1,1\}\to \{0,0,1,1,1\}}\to\frac{\widehat{C}_{\{1,1,1,1,1\}\to \{0,0,1,1,1\}}^{(2),N}}{\widehat{C}_{\{1,1,1,1,1\}\to \{0,0,1,1,1\}}^{(2),D}}\nn
  &&\longline{\widehat{C}_{\{1,1,1,1,1\}\to \{0,0,1,1,1\}}^{(2),N}}{s_{00} (s_{12} (s_{33}-s_{23})+s_{13} (s_{22}-s_{23})-s_{22} s_{33}+s_{23}^2)}\nn
  &&\longline{\widehat{C}_{\{1,1,1,1,1\}\to \{0,0,1,1,1\}}^{(2),D}}{s_{11} (s_{23}^2-s_{22} s_{33})-2 s_{12} s_{13} s_{23}+s_{12}^2 s_{33}+s_{13}^2 s_{22}}
  \nn
  &&\widehat{C}_{\{1,1,1,1,1\}\to \{1,0,1,1,0\}}\to{0}\nn  
  &&\partial_{M_0^2}C_{\{1,1,1,1\}\to \{1,0,1,1\}}\to\frac{\partial_{M_0^2}C_{\{1,1,1,1\}\to \{1,0,1,1\}}^{(2),N}}{\partial_{M_0^2}C_{\{1,1,1,1\}\to \{1,0,1,1\}}^{(2),D}}\nn
  &&\longline{\partial_{M_0^2}C_{\{1,1,1,1\}\to \{1,0,1,1\}}^{(2),N}}{s_{00} (s_{12} (s_{33}-s_{23})+s_{13} (s_{22}-s_{23})-s_{22} s_{33}+s_{23}^2)}\nn
  &&\longline{\partial_{M_0^2}C_{\{1,1,1,1\}\to \{1,0,1,1\}}^{(2),D}}{(D-3) (s_{11} (s_{23}^2-s_{22} s_{33})-2 s_{12} s_{13} s_{23}+s_{12}^2 s_{33}+s_{13}^2 s_{22})}\nonumber
  \eea 
  The coefficients of $I_{1, 1, 0, 1}$ can be obtained by the permutation $M_1 \leftrightarrow  M_2 $,$K_1 \leftrightarrow  K_2 $ on $I_{1, 0, 1, 1}$. Similarly, doing transformation $M_1 \leftrightarrow  M_3 $,$K_1 \leftrightarrow  K_3 $ one can get the coefficient of $I_{1, 1, 1, 0}$.
  
  \item {\bf $I_{1, 1, 1, 1}$}: The coresponding coefficients are
  \bea 
  &&\widehat{C}_{\{1,1,1,1,1\}\to \{0,1,1,1,1\}}\to\frac{\widehat{C}_{\{1,1,1,1,1\}\to \{0,1,1,1,1\}}^{(2),N}}{\widehat{C}_{\{1,1,1,1,1\}\to \{0,1,1,1,1\}}^{(2),D}}\nn
  &&\longline{\widehat{C}_{\{1,1,1,1,1\}\to \{0,1,1,1,1\}}^{(2),N}}{s_{00} (s_{11} (s_{12} (s_{23}-s_{33})+s_{13} (s_{23}-s_{22})-s_{22} s_{23}+M_0^2 s_{22}-M_3^2 s_{22}-s_{23} s_{33}-2 M_0^2 s_{23}+M_2^2 s_{23}+M_3^2 s_{23}+2 s_{23}^2+M_0^2 s_{33}-M_2^2 s_{33})+s_{12} (s_{13} (s_{22}-6 s_{23}+s_{33}+2 M_0^2-M_2^2-M_3^2)-s_{22} s_{33}+s_{23} s_{33}+2 M_0^2 s_{23}-M_1^2 s_{23}-M_3^2 s_{23}-2 M_0^2 s_{33}+M_1^2 s_{33}+M_2^2 s_{33})+s_{12}^2 (2 s_{33}-M_0^2+M_3^2)+s_{13} s_{22} s_{23}-s_{13} s_{22} s_{33}-2 M_0^2 s_{13} s_{22}+M_1^2 s_{13} s_{22}+M_3^2 s_{13} s_{22}+2 s_{13}^2 s_{22}+2 M_0^2 s_{13} s_{23}-M_1^2 s_{13} s_{23}-M_2^2 s_{13} s_{23}+M_0^2 (-s_{13}^2)+M_2^2 s_{13}^2+M_0^2 s_{22} s_{33}-M_1^2 s_{22} s_{33}-M_0^2 s_{23}^2+M_1^2 s_{23}^2)}\nn
  &&\longline{\widehat{C}_{\{1,1,1,1,1\}\to \{0,1,1,1,1\}}^{(2),D}}{s_{11} (s_{23}^2-s_{22} s_{33})-2 s_{12} s_{13} s_{23}+s_{12}^2 s_{33}+s_{13}^2 s_{22}}\nn
  &&\widehat{C}_{\{1,1,1,1,1\}\to \{1,1,1,1,0\}}\to {-s_{00}}\nn  
  &&\partial_{M_0^2}C_{\{1,1,1,1\}\to \{1,1,1,1\}}\to\frac{\partial_{M_0^2}C_{\{1,1,1,1\}\to \{1,1,1,1\}}^{(2),N}}{\partial_{M_0^2}C_{\{1,1,1,1\}\to \{1,1,1,1\}}^{(2),D}}\nn
  &&\longline{\partial_{M_0^2}C_{\{1,1,1,1\}\to \{1,1,1,1\}}^{(2),N}}{2 s_{00} (s_{11} (s_{12} (s_{23}-s_{33})+s_{13} (s_{23}-s_{22})-s_{22} s_{23}+s_{22} s_{33}+M_0^2 s_{22}-M_3^2 s_{22}-s_{23} s_{33}-2 M_0^2 s_{23}+M_2^2 s_{23}+M_3^2 s_{23}+s_{23}^2+M_0^2 s_{33}-M_2^2 s_{33})+s_{12} (s_{13} (s_{22}-4 s_{23}+s_{33}+2 M_0^2-M_2^2-M_3^2)-s_{22} s_{33}+s_{23} s_{33}+2 M_0^2 s_{23}-M_1^2 s_{23}-M_3^2 s_{23}-2 M_0^2 s_{33}+M_1^2 s_{33}+M_2^2 s_{33})+s_{12}^2 (s_{33}-M_0^2+M_3^2)+s_{13} s_{22} s_{23}-s_{13} s_{22} s_{33}-2 M_0^2 s_{13} s_{22}+M_1^2 s_{13} s_{22}+M_3^2 s_{13} s_{22}+s_{13}^2 s_{22}+2 M_0^2 s_{13} s_{23}-M_1^2 s_{13} s_{23}-M_2^2 s_{13} s_{23}+M_0^2 (-s_{13}^2)+M_2^2 s_{13}^2+M_0^2 s_{22} s_{33}-M_1^2 s_{22} s_{33}-M_0^2 s_{23}^2+M_1^2 s_{23}^2)}\nn
  &&\longline{\partial_{M_0^2}C_{\{1,1,1,1\}\to \{1,1,1,1\}}^{(2),D}}{(D-3) (s_{11} (s_{23}^2-s_{22} s_{33})-2 s_{12} s_{13} s_{23}+s_{12}^2 s_{33}+s_{13}^2 s_{22})}\nonumber 
  \ed\eea 
   
  \item {\bf $I_{2, 0, 0, 0}$}: The coresponding coefficients are
  \bea 
  &&\widehat{C}_{\{1,1,1,1,1\}\to \{1,0,0,0,1\}}\to 0,~~~~{C}_{\{1,1,1,1\}\to \{1,0,0,0\}}\to 0
  \ed\eea 
   
  \item {\bf $I_{2, 0, 0, 1}$}: The coresponding coefficients are
  \bea 
  &&\widehat{C}_{\{1,1,1,1,1\}\to \{1,0,0,1,1\}}\to 0,~~~~{C}_{\{1,1,1,1\}\to \{1,0,0,1\}}\to 0
  \ed\eea 
  According to permutation symmetry, coefficients of $I_{2, 0, 1, 0}$, $I_{2, 1, 0, 0}$ are both 0.    
     
  \item {\bf $I_{2, 0, 1, 1}$}: The coresponding coefficients are
  \bea 
  &&\widehat{C}_{\{1,1,1,1,1\}\to \{1,0,1,1,1\}}\to 0 \nn
  &&{C}_{\{1,1,1,1\}\to \{1,0,1,1\}}\to\frac{{C}_{\{1,1,1,1\}\to \{1,0,1,1\}}^{(2),N}}{{C}_{\{1,1,1,1\}\to \{1,0,1,1\}}^{(2),D}}\\
  &&\longline{{C}_{\{1,1,1,1\}\to \{1,0,1,1\}}^{(2),N}}{s_{00} ((s_{11}+M_0^2-M_1^2) (s_{23}^2-s_{22} s_{33})+(s_{33}+M_0^2-M_3^2) (s_{13} s_{22}-s_{12} s_{23})-(s_{22}+M_0^2-M_2^2) (s_{13} s_{23}-s_{12} s_{33}))}\nn
  &&\longline{{C}_{\{1,1,1,1\}\to \{1,0,1,1\}}^{(2),D}}{(D-3) (s_{11} (s_{23}^2-s_{22} s_{33})-2 s_{12} s_{13} s_{23}+s_{12}^2 s_{33}+s_{13}^2 s_{22})}\nonumber
  \ed\eea   
  The coefficients of other similar basis can be obtained by permutation symmetry.  
\end{itemize}
When putting all coefficients into \eref{frame-3-5}  we get the full results as given in the attached Mathematica file. 
%

\section{The pentagon}
First we have 
\begin{subequations}
	\allowdisplaybreaks
\bea
 & &\longline{\frac{1}{AR^2}}{\frac{1}{s_{00} \a_{2,1,1,1,1}}(D-4) (s_{34}^2 s_{12}^2-s_{33} s_{44} s_{12}^2+2 s_{13} (s_{23} s_{44}-s_{24} s_{34}) s_{12}-s_{11} s_{22} s_{34}^2-s_{11} s_{24}^2 s_{33}+s_{14}^2 (s_{23}^2-s_{22} s_{33})+2 s_{11} s_{23} s_{24} s_{34}-2 s_{14} (s_{12} (s_{23} s_{34}-s_{24} s_{33})+s_{13} (s_{23} s_{24}-s_{22} s_{34}))-s_{11} s_{23}^2 s_{44}+s_{11} s_{22} s_{33} s_{44}+s_{13}^2 (s_{24}^2-s_{22} s_{44}))}\eea
 where 
 \bea
 & &\longline{\a_{2,1,1,1,1}}{(-4 (s_{34}^2 s_{12}^2-s_{33} s_{44} s_{12}^2+2 s_{13} (s_{23} s_{44}-s_{24} s_{34}) s_{12}-s_{11} s_{22} s_{34}^2-s_{11} s_{24}^2 s_{33}+s_{14}^2 (s_{23}^2-s_{22} s_{33})+2 s_{11} s_{23} s_{24} s_{34}-2 s_{14} (s_{12} (s_{23} s_{34}-s_{24} s_{33})+s_{13} (s_{23} s_{24}-s_{22} s_{34}))-s_{11} s_{23}^2 s_{44}+s_{11} s_{22} s_{33} s_{44}+s_{13}^2 (s_{24}^2-s_{22} s_{44})) M_0^2+(M_0^2-M_4^2+s_{44}) ((M_0^2-M_2^2+s_{22}) (s_{24} s_{13}^2-(s_{14} s_{23}+s_{12} s_{34}) s_{13}+s_{12} s_{14} s_{33}+s_{11} (s_{23} s_{34}-s_{24} s_{33}))+(M_0^2-M_3^2+s_{33}) (s_{34} s_{12}^2-s_{12} s_{14} s_{23}+s_{13} (s_{14} s_{22}-s_{12} s_{24})+s_{11} (s_{23} s_{24}-s_{22} s_{34}))+(M_0^2-M_1^2+s_{11}) (s_{14} (s_{23}^2-s_{22} s_{33})+s_{13} (s_{22} s_{34}-s_{23} s_{24})+s_{12} (s_{24} s_{33}-s_{23} s_{34}))-(s_{33} s_{12}^2-2 s_{13} s_{23} s_{12}+s_{13}^2 s_{22}+s_{11} (s_{23}^2-s_{22} s_{33})) (M_0^2-M_4^2+s_{44}))+(M_0^2-M_3^2+s_{33}) ((s_{34} s_{12}^2-s_{12} s_{14} s_{23}+s_{13} (s_{14} s_{22}-s_{12} s_{24})+s_{11} (s_{23} s_{24}-s_{22} s_{34})) (M_0^2-M_4^2+s_{44})-(M_0^2-M_3^2+s_{33}) (s_{44} s_{12}^2-2 s_{14} s_{24} s_{12}+s_{14}^2 s_{22}+s_{11} (s_{24}^2-s_{22} s_{44}))+(M_0^2-M_2^2+s_{22}) (s_{23} s_{14}^2-(s_{13} s_{24}+s_{12} s_{34}) s_{14}+s_{12} s_{13} s_{44}+s_{11} (s_{24} s_{34}-s_{23} s_{44}))-(M_0^2-M_1^2+s_{11}) (s_{14} (s_{23} s_{24}-s_{22} s_{34})+s_{13} (s_{22} s_{44}-s_{24}^2)+s_{12} (s_{24} s_{34}-s_{23} s_{44})))-(M_0^2-M_2^2+s_{22}) (((s_{14} s_{23}+s_{12} s_{34}) s_{13}-s_{13}^2 s_{24}-s_{12} s_{14} s_{33}+s_{11} (s_{24} s_{33}-s_{23} s_{34})) (M_0^2-M_4^2+s_{44})-(M_0^2-M_3^2+s_{33}) (s_{23} s_{14}^2-(s_{13} s_{24}+s_{12} s_{34}) s_{14}+s_{12} s_{13} s_{44}+s_{11} (s_{24} s_{34}-s_{23} s_{44}))+(M_0^2-M_2^2+s_{22}) (s_{44} s_{13}^2-2 s_{14} s_{34} s_{13}+s_{14}^2 s_{33}+s_{11} (s_{34}^2-s_{33} s_{44}))-(M_0^2-M_1^2+s_{11}) (s_{14} (s_{24} s_{33}-s_{23} s_{34})+s_{13} (s_{23} s_{44}-s_{24} s_{34})+s_{12} (s_{34}^2-s_{33} s_{44})))+(M_0^2-M_1^2+s_{11}) ((s_{14} (s_{23}^2-s_{22} s_{33})+s_{13} (s_{22} s_{34}-s_{23} s_{24})+s_{12} (s_{24} s_{33}-s_{23} s_{34})) (M_0^2-M_4^2+s_{44})-(M_0^2-M_3^2+s_{33}) (s_{14} (s_{23} s_{24}-s_{22} s_{34})+s_{13} (s_{22} s_{44}-s_{24}^2)+s_{12} (s_{24} s_{34}-s_{23} s_{44}))+(M_0^2-M_2^2+s_{22}) (s_{14} (s_{24} s_{33}-s_{23} s_{34})+s_{13} (s_{23} s_{44}-s_{24} s_{34})+s_{12} (s_{34}^2-s_{33} s_{44}))-(M_0^2-M_1^2+s_{11}) (s_{44} s_{23}^2-2 s_{24} s_{34} s_{23}+s_{24}^2 s_{33}+s_{22} (s_{34}^2-s_{33} s_{44}))))}
\ed\eea
\end{subequations}

For various terms: 
\begin{itemize}
  \item The coefficients of tadpoles, bubbles and triangles are all zero because staring from the rank-2 hexagon, one can not reach those three topologies by the reduction.
  \item {\bf $I_{0, 1, 1, 1, 1}$}: The coresponding coefficients are
  \bea 
  &&\widehat{C}_{\{1,1,1,1,1,1\}\to \{0,1,1,1,1,0\}}\to\frac{\widehat{C}_{\{1,1,1,1,1,1\}\to \{0,1,1,1,1,0\}}^{(2),N}}{\widehat{C}_{\{1,1,1,1,1,1\}\to \{0,1,1,1,1,0\}}^{(2),D}}\nn
  &&\longline{\widehat{C}_{\{1,1,1,1,1,1\}\to \{0,1,1,1,1,0\}}^{(2),N}}{s_{00} (-s_{11} s_{22} s_{33}+2 s_{11} s_{22} s_{34}-s_{11} s_{22} s_{44}-2 s_{11} s_{23} s_{24}-2 s_{11} s_{23} s_{34}+2 s_{11} s_{23} s_{44}+s_{11} s_{23}^2+2 s_{11} s_{24} s_{33}-2 s_{11} s_{24} s_{34}+s_{11} s_{24}^2-s_{11} s_{33} s_{44}+s_{11} s_{34}^2-2 s_{13} (s_{12} (s_{23}-s_{24}-s_{34}+s_{44})+s_{14} (s_{22}-s_{23}-s_{24}+s_{34})+s_{22} s_{34}-s_{22} s_{44}-s_{23} s_{24}+s_{23} s_{44}-s_{24} s_{34}+s_{24}^2)+2 s_{14} (s_{12} (s_{23}-s_{24}-s_{33}+s_{34})+s_{22} s_{33}-s_{22} s_{34}+s_{23} (s_{24}+s_{34})-s_{23}^2-s_{24} s_{33})+2 s_{12} s_{23} s_{34}-2 s_{12} s_{23} s_{44}-2 s_{12} s_{24} s_{33}+2 s_{12} s_{24} s_{34}+2 s_{12} s_{33} s_{44}+s_{12}^2 s_{33}-2 s_{12}^2 s_{34}-2 s_{12} s_{34}^2+s_{12}^2 s_{44}+s_{13}^2 (s_{22}-2 s_{24}+s_{44})+s_{14}^2 (s_{22}-2 s_{23}+s_{33})-s_{22} s_{33} s_{44}+s_{22} s_{34}^2-2 s_{23} s_{24} s_{34}+s_{23}^2 s_{44}+s_{24}^2 s_{33})}\nn
  &&\longline{\widehat{C}_{\{1,1,1,1,1,1\}\to \{0,1,1,1,1,0\}}^{(2),D}}{s_{11} s_{22} s_{33} s_{44}-s_{11} s_{22} s_{34}^2+2 s_{11} s_{23} s_{24} s_{34}-s_{11} s_{23}^2 s_{44}-s_{11} s_{24}^2 s_{33}-2 s_{14} (s_{12} (s_{23} s_{34}-s_{24} s_{33})+s_{13} (s_{23} s_{24}-s_{22} s_{34}))+2 s_{12} s_{13} (s_{23} s_{44}-s_{24} s_{34})-s_{12}^2 s_{33} s_{44}+s_{12}^2 s_{34}^2+s_{13}^2 (s_{24}^2-s_{22} s_{44})+s_{14}^2 (s_{23}^2-s_{22} s_{33})}\nn
  &&\partial_{M_0^2}C_{\{1,1,1,1,1\}\to \{0,1,1,1,1\}}\to\frac{\partial_{M_0^2}C_{\{1,1,1,1,1\}\to \{0,1,1,1,1\}}^{(2),N}}{\partial_{M_0^2}C_{\{1,1,1,1,1\}\to \{0,1,1,1,1\}}^{(2),D}}\\
  &&\longline{\partial_{M_0^2}C_{\{1,1,1,1,1\}\to \{0,1,1,1,1\}}^{(2),N}}{s_{00} (-s_{11} s_{22} s_{33}+2 s_{11} s_{22} s_{34}-s_{11} s_{22} s_{44}-2 s_{11} s_{23} s_{24}-2 s_{11} s_{23} s_{34}+2 s_{11} s_{23} s_{44}+s_{11} s_{23}^2+2 s_{11} s_{24} s_{33}-2 s_{11} s_{24} s_{34}+s_{11} s_{24}^2-s_{11} s_{33} s_{44}+s_{11} s_{34}^2-2 s_{13} (s_{12} (s_{23}-s_{24}-s_{34}+s_{44})+s_{14} (s_{22}-s_{23}-s_{24}+s_{34})+s_{22} s_{34}-s_{22} s_{44}-s_{23} s_{24}+s_{23} s_{44}-s_{24} s_{34}+s_{24}^2)+2 s_{14} (s_{12} (s_{23}-s_{24}-s_{33}+s_{34})+s_{22} s_{33}-s_{22} s_{34}+s_{23} (s_{24}+s_{34})-s_{23}^2-s_{24} s_{33})+2 s_{12} s_{23} s_{34}-2 s_{12} s_{23} s_{44}-2 s_{12} s_{24} s_{33}+2 s_{12} s_{24} s_{34}+2 s_{12} s_{33} s_{44}+s_{12}^2 s_{33}-2 s_{12}^2 s_{34}-2 s_{12} s_{34}^2+s_{12}^2 s_{44}+s_{13}^2 (s_{22}-2 s_{24}+s_{44})+s_{14}^2 (s_{22}-2 s_{23}+s_{33})-s_{22} s_{33} s_{44}+s_{22} s_{34}^2-2 s_{23} s_{24} s_{34}+s_{23}^2 s_{44}+s_{24}^2 s_{33})}\nn
  &&\longline{\partial_{M_0^2}C_{\{1,1,1,1,1\}\to \{0,1,1,1,1\}}^{(2),D}}{(D-4) (s_{11} s_{22} s_{33} s_{44}-s_{11} s_{22} s_{34}^2+2 s_{11} s_{23} s_{24} s_{34}-s_{11} s_{23}^2 s_{44}-s_{11} s_{24}^2 s_{33}-2 s_{14} (s_{12} (s_{23} s_{34}-s_{24} s_{33})+s_{13} (s_{23} s_{24}-s_{22} s_{34}))+2 s_{12} s_{13} (s_{23} s_{44}-s_{24} s_{34})-s_{12}^2 s_{33} s_{44}+s_{12}^2 s_{34}^2+s_{13}^2 (s_{24}^2-s_{22} s_{44})+s_{14}^2 (s_{23}^2-s_{22} s_{33}))}\nonumber
  \ed\eea 
  
  \item {\bf $I_{1, 0, 1, 1, 1}$}: The corresponding coefficients are
  \bea 
  &&\widehat{C}_{\{1,1,1,1,1,1\}\to \{0,0,1,1,1,1\}}\to\frac{\widehat{C}_{\{1,1,1,1,1,1\}\to \{0,0,1,1,1,1\}}^{(2),N}}{\widehat{C}_{\{1,1,1,1,1,1\}\to \{0,0,1,1,1,1\}}^{(2),D}}\nn
  &&\longline{\widehat{C}_{\{1,1,1,1,1,1\}\to \{0,0,1,1,1,1\}}^{(2),N}}{s_{00} (s_{12} s_{23} s_{34}-s_{12} s_{23} s_{44}-s_{12} s_{24} s_{33}+s_{12} s_{24} s_{34}+s_{12} s_{33} s_{44}-s_{12} s_{34}^2+s_{13} (s_{22} (s_{44}-s_{34})+s_{23} (s_{24}-s_{44})+s_{24} s_{34}-s_{24}^2)+s_{14} (s_{22} (s_{33}-s_{34})+s_{23} (s_{24}+s_{34})-s_{23}^2-s_{24} s_{33})-s_{22} s_{33} s_{44}+s_{22} s_{34}^2-2 s_{23} s_{24} s_{34}+s_{23}^2 s_{44}+s_{24}^2 s_{33})}\nn
  &&\longline{\widehat{C}_{\{1,1,1,1,1,1\}\to \{0,0,1,1,1,1\}}^{(2),D}}{-s_{11} s_{22} s_{33} s_{44}+s_{11} s_{22} s_{34}^2-2 s_{11} s_{23} s_{24} s_{34}+s_{11} s_{23}^2 s_{44}+s_{11} s_{24}^2 s_{33}+2 s_{14} (s_{12} (s_{23} s_{34}-s_{24} s_{33})+s_{13} (s_{23} s_{24}-s_{22} s_{34}))+2 s_{12} s_{13} (s_{24} s_{34}-s_{23} s_{44})+s_{12}^2 s_{33} s_{44}+s_{12}^2 (-s_{34}^2)+s_{13}^2 (s_{22} s_{44}-s_{24}^2)+s_{14}^2 (s_{22} s_{33}-s_{23}^2)}\nn
  &&\widehat{C}_{\{1,1,1,1,1,1\}\to \{1,0,1,1,1,0\}}\to 0
  \\
  &&\partial_{M_0^2}C_{\{1,1,1,1,1\}\to \{1,0,1,1,1\}}\to\frac{\partial_{M_0^2}C_{\{1,1,1,1,1\}\to \{1,0,1,1,1\}}^{(2),N}}{\partial_{M_0^2}C_{\{1,1,1,1,1\}\to \{1,0,1,1,1\}}^{(2),D}}\nn
  &&\longline{\partial_{M_0^2}C_{\{1,1,1,1,1\}\to \{1,0,1,1,1\}}^{(2),N}}{s_{00} (-s_{12} s_{23} s_{34}+s_{12} s_{23} s_{44}+s_{12} s_{24} s_{33}-s_{12} s_{24} s_{34}-s_{12} s_{33} s_{44}+s_{12} s_{34}^2+s_{13} (s_{22} (s_{34}-s_{44})+s_{23} (s_{44}-s_{24})-s_{24} s_{34}+s_{24}^2)+s_{14} (s_{22} (s_{34}-s_{33})-s_{23} (s_{24}+s_{34})+s_{23}^2+s_{24} s_{33})+s_{22} s_{33} s_{44}-s_{22} s_{34}^2+2 s_{23} s_{24} s_{34}+s_{23}^2 (-s_{44})-s_{24}^2 s_{33})}\nn
  &&\longline{\partial_{M_0^2}C_{\{1,1,1,1,1\}\to \{1,0,1,1,1\}}^{(2),D}}{(D-4) (s_{11} s_{22} s_{33} s_{44}-s_{11} s_{22} s_{34}^2+2 s_{11} s_{23} s_{24} s_{34}-s_{11} s_{23}^2 s_{44}-s_{11} s_{24}^2 s_{33}-2 s_{14} (s_{12} (s_{23} s_{34}-s_{24} s_{33})+s_{13} (s_{23} s_{24}-s_{22} s_{34}))+2 s_{12} s_{13} (s_{23} s_{44}-s_{24} s_{34})-s_{12}^2 s_{33} s_{44}+s_{12}^2 s_{34}^2+s_{13}^2 (s_{24}^2-s_{22} s_{44})+s_{14}^2 (s_{23}^2-s_{22} s_{33}))}\nonumber
  \ed\eea 
  Similarly,we could obtain coefficients of $I_{1, 1, 0, 1, 1}$, $I_{1, 1, 1, 0, 1}$, $I_{1, 1, 1, 1, 0}$, $I_{1, 1, 1, 0, 0}$  by permutation. 
  
  \item {\bf $I_{1, 1, 1, 1, 1}$}: The several coresponding coefficients are
 \bean
 \widehat{C}_{\{1,1,1,1,1,1\}\to \{1,1,1,1,1,0\}}\to {-s_{00}} \eean
  \bean 
  &&\widehat{C}_{\{1,1,1,1,1,1\}\to \{0,1,1,1,1,1\}}\to\frac{\widehat{C}_{\{1,1,1,1,1,1\}\to \{0,1,1,1,1,1\}}^{(2),N}}{\widehat{C}_{\{1,1,1,1,1,1\}\to \{0,1,1,1,1,1\}}^{(2),D}}\eean
  where
  \bean & & \widehat{C}_{\{1,1,1,1,1,1\}\to \{0,1,1,1,1,1\}}^{(2),D}=  -s_{11} s_{22} s_{33} s_{44}+s_{11} s_{22} s_{34}^2-2 s_{11} s_{23} s_{24} s_{34}\nn %
  & & +s_{11} s_{23}^2 s_{44}+s_{11} s_{24}^2 s_{33}+2 s_{14} (s_{12} (s_{23} s_{34}-s_{24} s_{33})+s_{13} (s_{23} s_{24}-s_{22} s_{34}))\nn %
  & & +2 s_{12} s_{13} (s_{24} s_{34}-s_{23} s_{44})+s_{12}^2 s_{33} s_{44}+s_{12}^2 (-s_{34}^2)+s_{13}^2 (s_{22} s_{44}-s_{24}^2)+s_{14}^2 (s_{22} s_{33}-s_{23}^2)
	\eean
and
  \bean
  &&\longlongline{\widehat{C}_{\{1,1,1,1,1,1\}\to \{0,1,1,1,1,1\}}^{(2),N}}{s_{00} (-s_{11} s_{23}^2 M_0^2-s_{11} s_{24}^2 M_0^2-s_{11} s_{34}^2 M_0^2+2 s_{12} s_{34}^2 M_0^2-s_{22} s_{34}^2 M_0^2+2 s_{11} s_{23} s_{24} M_0^2-s_{12}^2 s_{33} M_0^2-s_{24}^2 s_{33} M_0^2+s_{11} s_{22} s_{33} M_0^2-2 s_{11} s_{24} s_{33} M_0^2+2 s_{12} s_{24} s_{33} M_0^2+2 s_{12}^2 s_{34} M_0^2-2 s_{11} s_{22} s_{34} M_0^2+2 s_{11} s_{23} s_{34} M_0^2-2 s_{12} s_{23} s_{34} M_0^2+2 s_{11} s_{24} s_{34} M_0^2-2 s_{12} s_{24} s_{34} M_0^2+2 s_{23} s_{24} s_{34} M_0^2-s_{12}^2 s_{44} M_0^2-s_{23}^2 s_{44} M_0^2+s_{11} s_{22} s_{44} M_0^2-2 s_{11} s_{23} s_{44} M_0^2+2 s_{12} s_{23} s_{44} M_0^2+s_{11} s_{33} s_{44} M_0^2-2 s_{12} s_{33} s_{44} M_0^2+s_{22} s_{33} s_{44} M_0^2+M_4^2 s_{11} s_{23}^2+M_3^2 s_{11} s_{24}^2-3 s_{12}^2 s_{34}^2+M_2^2 s_{11} s_{34}^2-M_1^2 s_{12} s_{34}^2-M_2^2 s_{12} s_{34}^2+s_{11} s_{12} s_{34}^2+M_1^2 s_{22} s_{34}^2+s_{11} s_{22} s_{34}^2+s_{12} s_{22} s_{34}^2-M_3^2 s_{11} s_{23} s_{24}-M_4^2 s_{11} s_{23} s_{24}+M_4^2 s_{12}^2 s_{33}+M_1^2 s_{24}^2 s_{33}+s_{11} s_{24}^2 s_{33}-M_4^2 s_{11} s_{22} s_{33}+M_2^2 s_{11} s_{24} s_{33}+M_4^2 s_{11} s_{24} s_{33}-M_1^2 s_{12} s_{24} s_{33}-M_4^2 s_{12} s_{24} s_{33}+s_{11} s_{12} s_{24} s_{33}-s_{11} s_{22} s_{24} s_{33}+s_{11} s_{23} s_{24} s_{33}+s_{14}^2 (-3 s_{23}^2+(2 M_0^2-M_2^2-M_3^2+s_{33}) s_{23}+(M_2^2-M_0^2) s_{33}+s_{22} (-M_0^2+M_3^2+s_{23}+s_{33}))-M_3^2 s_{12}^2 s_{34}-M_4^2 s_{12}^2 s_{34}+M_3^2 s_{11} s_{22} s_{34}+M_4^2 s_{11} s_{22} s_{34}-M_2^2 s_{11} s_{23} s_{34}-M_4^2 s_{11} s_{23} s_{34}+M_1^2 s_{12} s_{23} s_{34}+M_4^2 s_{12} s_{23} s_{34}-s_{11} s_{12} s_{23} s_{34}+s_{11} s_{22} s_{23} s_{34}-M_2^2 s_{11} s_{24} s_{34}-M_3^2 s_{11} s_{24} s_{34}+M_1^2 s_{12} s_{24} s_{34}+M_3^2 s_{12} s_{24} s_{34}-s_{11} s_{12} s_{24} s_{34}+s_{11} s_{22} s_{24} s_{34}-2 M_1^2 s_{23} s_{24} s_{34}-4 s_{11} s_{23} s_{24} s_{34}+s_{12}^2 s_{33} s_{34}-s_{11} s_{22} s_{33} s_{34}+s_{11} s_{24} s_{33} s_{34}-s_{12} s_{24} s_{33} s_{34}+M_3^2 s_{12}^2 s_{44}+M_1^2 s_{23}^2 s_{44}+s_{11} s_{23}^2 s_{44}-M_3^2 s_{11} s_{22} s_{44}+M_2^2 s_{11} s_{23} s_{44}+M_3^2 s_{11} s_{23} s_{44}-M_1^2 s_{12} s_{23} s_{44}-M_3^2 s_{12} s_{23} s_{44}+s_{11} s_{12} s_{23} s_{44}-s_{11} s_{22} s_{23} s_{44}+s_{11} s_{23} s_{24} s_{44}+s_{12}^2 s_{33} s_{44}-M_2^2 s_{11} s_{33} s_{44}+M_1^2 s_{12} s_{33} s_{44}+M_2^2 s_{12} s_{33} s_{44}-s_{11} s_{12} s_{33} s_{44}-M_1^2 s_{22} s_{33} s_{44}+s_{11} s_{22} s_{33} s_{44}-s_{12} s_{22} s_{33} s_{44}-s_{11} s_{23} s_{33} s_{44}+s_{12} s_{23} s_{33} s_{44}-s_{11} s_{24} s_{33} s_{44}+s_{12} s_{24} s_{33} s_{44}+s_{12}^2 s_{34} s_{44}-s_{11} s_{22} s_{34} s_{44}+s_{11} s_{23} s_{34} s_{44}-s_{12} s_{23} s_{34} s_{44}+s_{13}^2 (-3 s_{24}^2+(2 M_0^2-M_2^2-M_4^2+s_{44}) s_{24}+(M_2^2-M_0^2) s_{44}+s_{22} (-M_0^2+M_4^2+s_{24}+s_{44}))+s_{14} (2 s_{23}^2 M_0^2-2 s_{23} s_{24} M_0^2-2 s_{22} s_{33} M_0^2+2 s_{24} s_{33} M_0^2+2 s_{22} s_{34} M_0^2-2 s_{23} s_{34} M_0^2-M_1^2 s_{23}^2-M_4^2 s_{23}^2+M_1^2 s_{23} s_{24}+M_3^2 s_{23} s_{24}+M_1^2 s_{22} s_{33}+M_4^2 s_{22} s_{33}-M_1^2 s_{24} s_{33}-M_2^2 s_{24} s_{33}+s_{22} s_{24} s_{33}-s_{23} s_{24} s_{33}-M_1^2 s_{22} s_{34}-M_3^2 s_{22} s_{34}+M_1^2 s_{23} s_{34}+M_2^2 s_{23} s_{34}-s_{22} s_{23} s_{34}+s_{22} s_{33} s_{34}+s_{11} (s_{23}^2-(s_{24}+s_{34}) s_{23}+s_{24} s_{33}+s_{22} (s_{34}-s_{33}))+s_{23}^2 s_{44}-s_{22} s_{33} s_{44}+s_{12} (2 s_{24} M_0^2+2 s_{33} M_0^2-2 s_{34} M_0^2-2 M_3^2 s_{24}-M_2^2 s_{33}-M_4^2 s_{33}-4 s_{24} s_{33}+s_{22} (s_{33}-s_{34})+M_2^2 s_{34}+M_3^2 s_{34}-s_{33} s_{34}+s_{23} (-2 M_0^2+M_3^2+M_4^2-s_{33}+6 s_{34}-s_{44})+s_{33} s_{44}))-s_{13} (-2 s_{24}^2 M_0^2-2 s_{12} s_{23} M_0^2+2 s_{12} s_{24} M_0^2+2 s_{23} s_{24} M_0^2+2 s_{12} s_{34} M_0^2-2 s_{22} s_{34} M_0^2+2 s_{24} s_{34} M_0^2-2 s_{12} s_{44} M_0^2+2 s_{22} s_{44} M_0^2-2 s_{23} s_{44} M_0^2+M_1^2 s_{24}^2+M_3^2 s_{24}^2+2 M_4^2 s_{12} s_{23}-M_3^2 s_{12} s_{24}-M_4^2 s_{12} s_{24}-M_1^2 s_{23} s_{24}-M_4^2 s_{23} s_{24}-s_{24}^2 s_{33}+s_{12} s_{24} s_{33}-M_2^2 s_{12} s_{34}-M_4^2 s_{12} s_{34}+M_1^2 s_{22} s_{34}+M_4^2 s_{22} s_{34}+s_{12} s_{22} s_{34}-M_1^2 s_{24} s_{34}-M_2^2 s_{24} s_{34}-6 s_{12} s_{24} s_{34}+s_{22} s_{24} s_{34}+s_{14} (2 s_{24} M_0^2-2 s_{34} M_0^2-M_2^2 s_{24}-M_3^2 s_{24}+s_{24} s_{33}+2 M_2^2 s_{34}-s_{23} (-2 M_0^2+M_2^2+M_4^2+6 s_{24}-s_{44})+s_{22} (-2 M_0^2+M_3^2+M_4^2+s_{23}+s_{24}-s_{33}+4 s_{34}-s_{44}))+M_2^2 s_{12} s_{44}+M_3^2 s_{12} s_{44}-M_1^2 s_{22} s_{44}-M_3^2 s_{22} s_{44}-s_{12} s_{22} s_{44}+M_1^2 s_{23} s_{44}+M_2^2 s_{23} s_{44}+4 s_{12} s_{23} s_{44}-s_{22} s_{23} s_{44}+s_{12} s_{24} s_{44}+s_{23} s_{24} s_{44}-s_{12} s_{33} s_{44}+s_{22} s_{33} s_{44}+s_{12} s_{34} s_{44}-s_{22} s_{34} s_{44}+s_{11} (-s_{24}^2+s_{34} s_{24}+s_{23} (s_{24}-s_{44})+s_{22} (s_{44}-s_{34}))))}\eean
  \bea 
  &&\partial_{M_0^2}C_{\{1,1,1,1,1\}\to \{1,1,1,1,1\}}\to\frac{\partial_{M_0^2}C_{\{1,1,1,1,1\}\to \{1,1,1,1,1\}}^{(2),N}}{\partial_{M_0^2}C_{\{1,1,1,1,1\}\to \{1,1,1,1,1\}}^{(2),D}}\nn
  &&\longlongline{\partial_{M_0^2}C_{\{1,1,1,1,1\}\to \{1,1,1,1,1\}}^{(2),N}}{2 s_{00} (s_{11} s_{23}^2 M_0^2+s_{11} s_{24}^2 M_0^2+s_{11} s_{34}^2 M_0^2-2 s_{12} s_{34}^2 M_0^2+s_{22} s_{34}^2 M_0^2-2 s_{11} s_{23} s_{24} M_0^2+s_{12}^2 s_{33} M_0^2+s_{24}^2 s_{33} M_0^2-s_{11} s_{22} s_{33} M_0^2+2 s_{11} s_{24} s_{33} M_0^2-2 s_{12} s_{24} s_{33} M_0^2-2 s_{12}^2 s_{34} M_0^2+2 s_{11} s_{22} s_{34} M_0^2-2 s_{11} s_{23} s_{34} M_0^2+2 s_{12} s_{23} s_{34} M_0^2-2 s_{11} s_{24} s_{34} M_0^2+2 s_{12} s_{24} s_{34} M_0^2-2 s_{23} s_{24} s_{34} M_0^2+s_{12}^2 s_{44} M_0^2+s_{23}^2 s_{44} M_0^2-s_{11} s_{22} s_{44} M_0^2+2 s_{11} s_{23} s_{44} M_0^2-2 s_{12} s_{23} s_{44} M_0^2-s_{11} s_{33} s_{44} M_0^2+2 s_{12} s_{33} s_{44} M_0^2-s_{22} s_{33} s_{44} M_0^2-M_4^2 s_{11} s_{23}^2-M_3^2 s_{11} s_{24}^2+2 s_{12}^2 s_{34}^2-M_2^2 s_{11} s_{34}^2+M_1^2 s_{12} s_{34}^2+M_2^2 s_{12} s_{34}^2-s_{11} s_{12} s_{34}^2-M_1^2 s_{22} s_{34}^2-s_{12} s_{22} s_{34}^2+M_3^2 s_{11} s_{23} s_{24}+M_4^2 s_{11} s_{23} s_{24}-M_4^2 s_{12}^2 s_{33}-M_1^2 s_{24}^2 s_{33}+M_4^2 s_{11} s_{22} s_{33}-M_2^2 s_{11} s_{24} s_{33}-M_4^2 s_{11} s_{24} s_{33}+M_1^2 s_{12} s_{24} s_{33}+M_4^2 s_{12} s_{24} s_{33}-s_{11} s_{12} s_{24} s_{33}+s_{11} s_{22} s_{24} s_{33}-s_{11} s_{23} s_{24} s_{33}+s_{14}^2 (2 s_{23}^2+(-2 M_0^2+M_2^2+M_3^2-s_{33}) s_{23}-s_{22} (-M_0^2+M_3^2+s_{23})+(M_0^2-M_2^2) s_{33})+M_3^2 s_{12}^2 s_{34}+M_4^2 s_{12}^2 s_{34}-M_3^2 s_{11} s_{22} s_{34}-M_4^2 s_{11} s_{22} s_{34}+M_2^2 s_{11} s_{23} s_{34}+M_4^2 s_{11} s_{23} s_{34}-M_1^2 s_{12} s_{23} s_{34}-M_4^2 s_{12} s_{23} s_{34}+s_{11} s_{12} s_{23} s_{34}-s_{11} s_{22} s_{23} s_{34}+M_2^2 s_{11} s_{24} s_{34}+M_3^2 s_{11} s_{24} s_{34}-M_1^2 s_{12} s_{24} s_{34}-M_3^2 s_{12} s_{24} s_{34}+s_{11} s_{12} s_{24} s_{34}-s_{11} s_{22} s_{24} s_{34}+2 M_1^2 s_{23} s_{24} s_{34}+2 s_{11} s_{23} s_{24} s_{34}-s_{12}^2 s_{33} s_{34}+s_{11} s_{22} s_{33} s_{34}-s_{11} s_{24} s_{33} s_{34}+s_{12} s_{24} s_{33} s_{34}-M_3^2 s_{12}^2 s_{44}-M_1^2 s_{23}^2 s_{44}+M_3^2 s_{11} s_{22} s_{44}-M_2^2 s_{11} s_{23} s_{44}-M_3^2 s_{11} s_{23} s_{44}+M_1^2 s_{12} s_{23} s_{44}+M_3^2 s_{12} s_{23} s_{44}-s_{11} s_{12} s_{23} s_{44}+s_{11} s_{22} s_{23} s_{44}-s_{11} s_{23} s_{24} s_{44}+M_2^2 s_{11} s_{33} s_{44}-M_1^2 s_{12} s_{33} s_{44}-M_2^2 s_{12} s_{33} s_{44}+s_{11} s_{12} s_{33} s_{44}+M_1^2 s_{22} s_{33} s_{44}-2 s_{11} s_{22} s_{33} s_{44}+s_{12} s_{22} s_{33} s_{44}+s_{11} s_{23} s_{33} s_{44}-s_{12} s_{23} s_{33} s_{44}+s_{11} s_{24} s_{33} s_{44}-s_{12} s_{24} s_{33} s_{44}-s_{12}^2 s_{34} s_{44}+s_{11} s_{22} s_{34} s_{44}-s_{11} s_{23} s_{34} s_{44}+s_{12} s_{23} s_{34} s_{44}+s_{13}^2 (2 s_{24}^2+(-2 M_0^2+M_2^2+M_4^2-s_{44}) s_{24}-s_{22} (-M_0^2+M_4^2+s_{24})+(M_0^2-M_2^2) s_{44})+s_{14} (-2 s_{23}^2 M_0^2+2 s_{23} s_{24} M_0^2+2 s_{22} s_{33} M_0^2-2 s_{24} s_{33} M_0^2-2 s_{22} s_{34} M_0^2+2 s_{23} s_{34} M_0^2+M_1^2 s_{23}^2+M_4^2 s_{23}^2-M_1^2 s_{23} s_{24}-M_3^2 s_{23} s_{24}-M_1^2 s_{22} s_{33}-M_4^2 s_{22} s_{33}+M_1^2 s_{24} s_{33}+M_2^2 s_{24} s_{33}-s_{22} s_{24} s_{33}+s_{23} s_{24} s_{33}+M_1^2 s_{22} s_{34}+M_3^2 s_{22} s_{34}-M_1^2 s_{23} s_{34}-M_2^2 s_{23} s_{34}+s_{22} s_{23} s_{34}-s_{22} s_{33} s_{34}+s_{11} (-s_{23}^2+(s_{24}+s_{34}) s_{23}-s_{24} s_{33}+s_{22} (s_{33}-s_{34}))-s_{23}^2 s_{44}+s_{22} s_{33} s_{44}+s_{12} (-2 s_{24} M_0^2-2 s_{33} M_0^2+2 s_{34} M_0^2+2 M_3^2 s_{24}+M_2^2 s_{33}+M_4^2 s_{33}+2 s_{24} s_{33}-M_2^2 s_{34}-M_3^2 s_{34}+s_{33} s_{34}+s_{22} (s_{34}-s_{33})-s_{33} s_{44}+s_{23} (2 M_0^2-M_3^2-M_4^2+s_{33}-4 s_{34}+s_{44})))+s_{13} (-2 s_{24}^2 M_0^2-2 s_{12} s_{23} M_0^2+2 s_{12} s_{24} M_0^2+2 s_{23} s_{24} M_0^2+2 s_{12} s_{34} M_0^2-2 s_{22} s_{34} M_0^2+2 s_{24} s_{34} M_0^2-2 s_{12} s_{44} M_0^2+2 s_{22} s_{44} M_0^2-2 s_{23} s_{44} M_0^2+M_1^2 s_{24}^2+M_3^2 s_{24}^2+2 M_4^2 s_{12} s_{23}-M_3^2 s_{12} s_{24}-M_4^2 s_{12} s_{24}-M_1^2 s_{23} s_{24}-M_4^2 s_{23} s_{24}-s_{24}^2 s_{33}+s_{12} s_{24} s_{33}-M_2^2 s_{12} s_{34}-M_4^2 s_{12} s_{34}+M_1^2 s_{22} s_{34}+M_4^2 s_{22} s_{34}+s_{12} s_{22} s_{34}-M_1^2 s_{24} s_{34}-M_2^2 s_{24} s_{34}-4 s_{12} s_{24} s_{34}+s_{22} s_{24} s_{34}+s_{14} (2 s_{24} M_0^2-2 s_{34} M_0^2-M_2^2 s_{24}-M_3^2 s_{24}+s_{24} s_{33}+2 M_2^2 s_{34}-s_{23} (-2 M_0^2+M_2^2+M_4^2+4 s_{24}-s_{44})+s_{22} (-2 M_0^2+M_3^2+M_4^2+s_{23}+s_{24}-s_{33}+2 s_{34}-s_{44}))+M_2^2 s_{12} s_{44}+M_3^2 s_{12} s_{44}-M_1^2 s_{22} s_{44}-M_3^2 s_{22} s_{44}-s_{12} s_{22} s_{44}+M_1^2 s_{23} s_{44}+M_2^2 s_{23} s_{44}+2 s_{12} s_{23} s_{44}-s_{22} s_{23} s_{44}+s_{12} s_{24} s_{44}+s_{23} s_{24} s_{44}-s_{12} s_{33} s_{44}+s_{22} s_{33} s_{44}+s_{12} s_{34} s_{44}-s_{22} s_{34} s_{44}+s_{11} (-s_{24}^2+s_{34} s_{24}+s_{23} (s_{24}-s_{44})+s_{22} (s_{44}-s_{34}))))}\nn
  &&\longline{\partial_{M_0^2}C_{\{1,1,1,1,1\}\to \{1,1,1,1,1\}}^{(2),D}}{(D-4) (s_{11} s_{22} s_{33} s_{44}-s_{11} s_{22} s_{34}^2+2 s_{11} s_{23} s_{24} s_{34}-s_{11} s_{23}^2 s_{44}-s_{11} s_{24}^2 s_{33}-2 s_{14} (s_{12} (s_{23} s_{34}-s_{24} s_{33})+s_{13} (s_{23} s_{24}-s_{22} s_{34}))+2 s_{12} s_{13} (s_{23} s_{44}-s_{24} s_{34})-s_{12}^2 s_{33} s_{44}+s_{12}^2 s_{34}^2+s_{13}^2 (s_{24}^2-s_{22} s_{44})+s_{14}^2 (s_{23}^2-s_{22} s_{33}))}\nonumber
  \eea 
  
  \item {\bf $I_{2, 0, 0, 0, 0}$}: The coresponding coefficients are
  \bea 
  &&\widehat{C}_{\{1,1,1,1,1,1\}\to \{1,0,0,0,0,1\}}\to 0 ,~~~
  {C}_{\{1,1,1,1,1\}\to \{1,0,0,0,0\}}\to 0
  \eea 
  
  \item {\bf $I_{2, 0, 0, 0, 1}$}: The coresponding coefficients are
  \bea 
  &&\widehat{C}_{\{1,1,1,1,1,1\}\to \{1,0,0,0,1,1\}}\to 0 ,~~~
  {C}_{\{1,1,1,1,1\}\to \{1,0,0,0,1\}}\to 0
  \eea 
    
  \item {\bf $I_{2, 0, 0, 1, 1}$}: The coresponding coefficients are
  \bea 
  &&\widehat{C}_{\{1,1,1,1,1,1\}\to \{1,0,0,1,1,1\}}\to 0 ,~~~
  {C}_{\{1,1,1,1,1\}\to \{1,0,0,1,1\}}\to 0
  \eea 
  
  \item {\bf $I_{2, 0, 1, 1, 1}$}: The corresponding coefficients are
  \bea 
  &&\widehat{C}_{\{1,1,1,1,1,1\}\to \{1,0,1,1,1,1\}}\to{0}\nn  
  &&{C}_{\{1,1,1,1,1\}\to \{1,0,1,1,1\}}\to\frac{{C}_{\{1,1,1,1,1\}\to \{1,0,1,1,1\}}^{(2),N}}{{C}_{\{1,1,1,1,1\}\to \{1,0,1,1,1\}}^{(2),D}}\nn
  &&\longline{{C}_{\{1,1,1,1,1\}\to \{1,0,1,1,1\}}^{(2),N}}{-s_{00} ((s_{11}+M_0^2-M_1^2) (s_{22} (s_{34}^2-s_{33} s_{44})-2 s_{23} s_{24} s_{34}+s_{23}^2 s_{44}+s_{24}^2 s_{33})-((s_{44}+M_0^2-M_4^2) (s_{12} (s_{24} s_{33}-s_{23} s_{34})+s_{13} (s_{22} s_{34}-s_{23} s_{24})+s_{14} (s_{23}^2-s_{22} s_{33})))+(s_{33}+M_0^2-M_3^2) (s_{12} (s_{24} s_{34}-s_{23} s_{44})+s_{13} (s_{22} s_{44}-s_{24}^2)+s_{14} (s_{23} s_{24}-s_{22} s_{34}))-(s_{22}+M_0^2-M_2^2) (s_{12} (s_{34}^2-s_{33} s_{44})+s_{13} (s_{23} s_{44}-s_{24} s_{34})+s_{14} (s_{24} s_{33}-s_{23} s_{34})))}\nn
  &&\longline{{C}_{\{1,1,1,1,1\}\to \{1,0,1,1,1\}}^{(2),D}}{(D-4) (s_{11} s_{22} s_{33} s_{44}-s_{11} s_{22} s_{34}^2+2 s_{11} s_{23} s_{24} s_{34}-s_{11} s_{23}^2 s_{44}-s_{11} s_{24}^2 s_{33}-2 s_{14} (s_{12} (s_{23} s_{34}-s_{24} s_{33})+s_{13} (s_{23} s_{24}-s_{22} s_{34}))+2 s_{12} s_{13} (s_{23} s_{44}-s_{24} s_{34})-s_{12}^2 s_{33} s_{44}+s_{12}^2 s_{34}^2+s_{13}^2 (s_{24}^2-s_{22} s_{44})+s_{14}^2 (s_{23}^2-s_{22} s_{33}))}
  \eea      
\end{itemize}


\bibliographystyle{JHEP}
\bibliography{reduction}

\end{document}